\documentclass[aps,prb,twocolumn,superscriptaddress,floatfix]{revtex4-1}
\pdfoutput=1
\usepackage{amsmath,amssymb}
\usepackage{graphicx}
\usepackage{color}
\usepackage{bm}
\usepackage{textcomp} 
\usepackage{enumitem}
\usepackage{gensymb}
\usepackage{xcolor}
\usepackage{hyperref}
\usepackage{algorithm,algpseudocode}
\usepackage{listings}
\usepackage{placeins}
\hypersetup{
    colorlinks=true,
    linkcolor=blue,
    urlcolor=blue,
    citecolor=blue,
}
\setlist{noitemsep,leftmargin=*}

\def\tbf{\textbf}

\begin{document}

\title{Dynamic Compressed Sensing for Real-Time Tomographic Reconstruction}

\author{Jonathan Schwartz}
\affiliation{Department of Material Science and Engineering, Ann Arbor,University of Michigan, MI, USA}
\author{Huihuo Zheng}
\affiliation{Argonne Leadership Computing Facility, Argonne National Laboratory, Lemont, IL, USA}
\author{Marcus Hanwell}
\affiliation{Kitware Inc, Clifton Park, NY, USA}
\author{Yi Jiang}
\affiliation{Advanced Photon Source Facility, Argonne National Laboratory, Lemont, IL, USA}
\author{Robert Hovden}
\email[Corresponding Author: ]{hovden@umich.edu}
\affiliation{Department of Material Science and Engineering, Ann Arbor,University of Michigan, MI, USA}
\affiliation{Applied Physics Program, University of Michigan, Ann Arbor, MI, USA}

\begin{abstract}
\textsc{\bf Abstract}: Electron tomography has achieved higher resolution and quality at reduced doses with recent advances in compressed sensing. Compressed sensing (CS) theory exploits the inherent sparse signal structure to efficiently reconstruct three-dimensional (3D) volumes at the nanoscale from undersampled measurements. However, the process bottlenecks 3D reconstruction with computation times that run from hours to days. Here we demonstrate a framework for dynamic compressed sensing that produces a 3D specimen structure that updates in real-time as new specimen projections are collected. Researchers can begin interpreting 3D specimens as data is collected to facilitate high-throughput and interactive analysis. Using scanning transmission electron microscopy (STEM), we show that dynamic compressed sensing accelerates the convergence speed by $\sim$3-fold while also reducing its error by 27\% for a Au/SrTiO$_3$ nanoparticle specimen. Before a tomography experiment is completed, the 3D tomogram has interpretable structure within $\sim$33\% of completion and fine details are visible as early as $\sim$66\%. Upon completion of an experiment, a high-fidelity 3D visualization is produced without further delay. Additionally, reconstruction parameters that tune data fidelity can be manipulated throughout the computation without rerunning the entire process.
\end{abstract}

\maketitle

\section{Introduction}
Electron tomography can reveal three-dimensional (3D) structure of biological specimens \cite{klug1968et,2010cryoem} and inorganic materials \cite{yu2012fuelcell,sai2013copolymer,ercius2009cu} across the nano- to meso- scale. However, 3D reconstruction commonly suffers from an insufficient number of projections and a limited angular range inside the microscope, commonly referred to as the missing wedge \cite{yi2017cslimits}. To overcome these experimental limitations, advancements in tomographic reconstruction algorithms now utilize compressed sensing (CS) methods that leverage the notion of sparsity to provide superior 3D resolution with limited angles and at lower doses \cite{donoho2006cs,candes2006cs,midgley2011csnp}. While compressed sensing algorithms provide higher quality reconstructions, they require substantially more computation time to complete and bottleneck electron tomography. Even worse, incomplete convergence and poor selection of tunable parameters are often discovered at the end of an arduous reconstruction---multiplying the total time required to produce a full 3D tomogram. Scientists would benefit from immediate feedback to examine intermediate results and optimize experimental parameters that accelerate the end-to-end scientific process \cite{bicer2015alcftomo}. 





Here, we have developed a dynamic CS framework that offers a 3D specimen reconstruction in real-time as projection data is collected. It enables direct feedback and on the fly optimization of experimental parameters. The reconstruction algorithm begins immediately upon acquiring the first projection and dynamically updates the 3D structure as new projections arrive---unlike traditional schemes which start after the experiment is complete. This means researchers can start analysis and characterization with high-fidelity tomograms before an experiment is complete. Using scanning transmission electron microscope (STEM) tomography \cite{midgley2003ztomo}, we demonstrate our method accelerates the final convergence by a factor of 2-3 over conventional CS and provides insight into 3D nanostructure within 62\% of the total experimental acquisition time. Dynamic reconstruction reduced the reconstruction error for Au/SrTiO$_3$ nanoparticles by 27\% and converged 100\% faster than a traditional approach. Moreover, researchers can use dynamic CS to manipulate the data-tolerance throughout the reconstruction and efficiently explore tunable parameters without having completely reset the algorithm. Implementing dynamic CS requires complete parallelization that includes the 3D total variation regularization for the isotropic norm. Tomograms ($\sim512^3~\text{voxels}$) can reconstructed dynamically on modest multi-core laptops during an electron tomography experiment and larger reconstructions ($\sim2048^3$) are achievable with high performance computing.

\section{Background}

The notion of sparsity has become widely used in signal processing and image reconstruction as a prior knowledge to regularize solutions in underdetermined problems. It was greatly popularized by the theory of compressed sensing \cite{sidky2011constrainedtv} that demonstrates the possibility to accurately recover the 3D structure of specimens ($\hat{\tbf{x}}$) from an insufficient number of projections ($b$) with $\ell_1$-norm optimizations. One of the most representative sparsity-exploiting algorithms is the total variation minimization (TV-min), which was originally proposed for image denoising \cite{osher1992tv} and widely used to reduce tomographic artifacts for reconstructions from a limited number of projections \cite{sidky2006tomotv,midgley2013cs,guay2016csbio}. The technique can effectively remove noisy features while preserving the edges of the object by minimizing its gradient magnitude. In this work, we consider a constrained optimization problem defined as: 

\begin{align}
    \hat{\tbf{x}}^* = \text{arg}\min_{\hat{\tbf{x}} \geq 0} \|\hat{\tbf{x}} \|_{\text{TV}} \text{~s.t.~} \|A \hat{\tbf{x}}-b\|_2 \leq \epsilon \label{eq:1}
\end{align} 
where $\|.\|_{\text{TV}}$ and $\|.\|_2$ denote the TV and $\ell_2$ norms, and $\epsilon$ is a data-tolerance parameter that controls the trade-off between regularization (smoothness) and data fidelity. Here, a tomography experiment is formulated as an inverse problem, $A\hat{\tbf{x}}=b$, where $A$ is the projection matrix that models the measurement physics. The constrained optimization problem can be solved with a combination of adaptive steepest-descent (ASD) to minimize TV and projection onto convex sets (POCS) to enforce data constraints---commonly referred to as (ASD-POCS) \cite{sidky2008asd}. This constrained optimization provides physical meaning to the tunable regularization parameter, $\epsilon$, which can be initially estimated from the data quality \cite{liu2017alttomo}. Across all `flavors' of compressed sensing tomography, the optimization process begins after all data has been collected. The iterative process can take thousands of iterations and runs from hours to a full day before converging to the designed solution. Moreover, the regularization parameter (here, $\epsilon$) is often task-dependent and needs be to adjusted to produce the best image quality, further increasing the computation time for the reconstruction process.

\begin{figure}[ht]
    \includegraphics[width=\columnwidth]{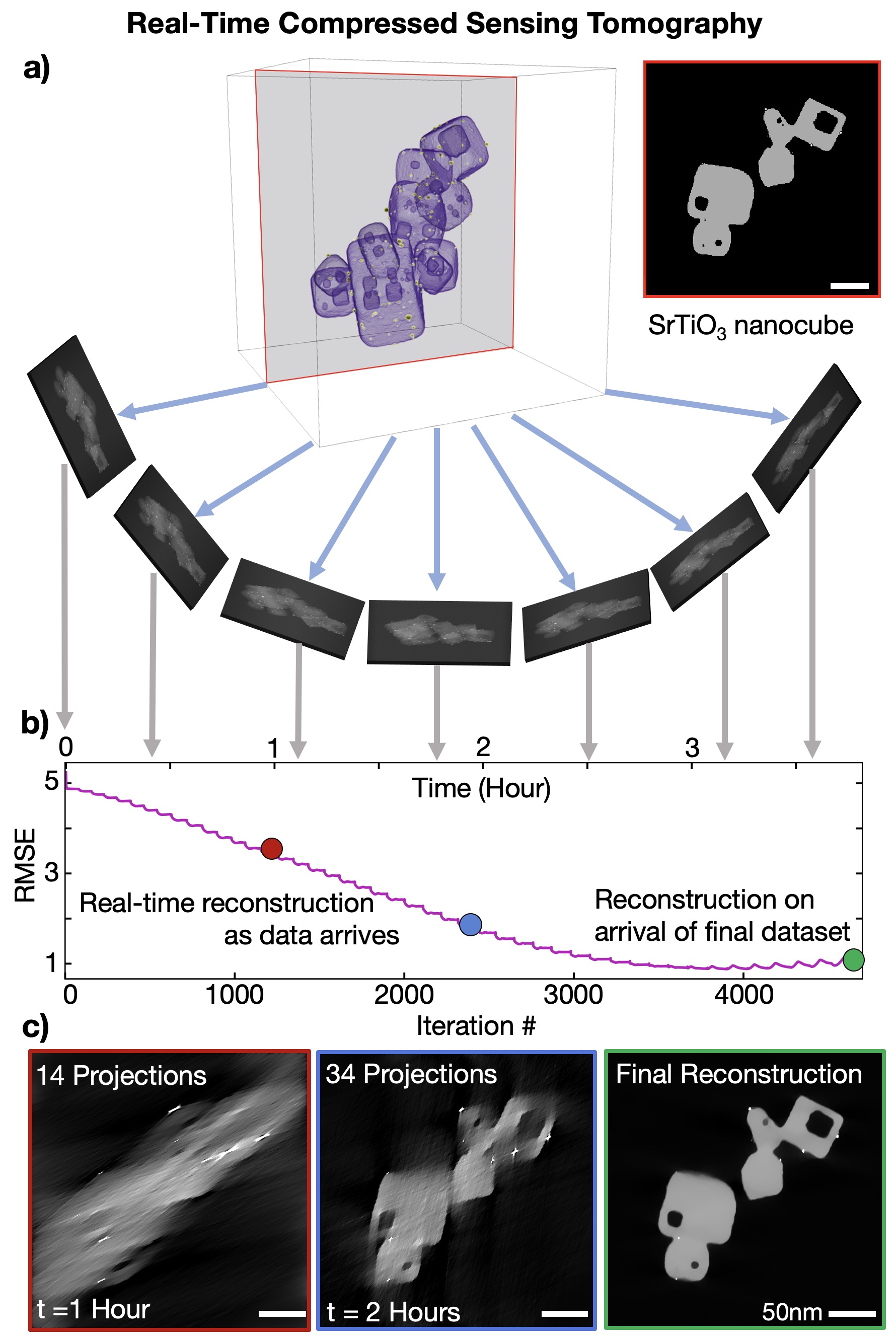}
    \caption{\textbf{Overview of the dynamic CS framework.} a) As the tomographic experiment progresses, projections are collected across an angular range (typically $\pm 75\degree$ for electron tomography and $\pm 90\degree$ in X-ray tomography). Measured projections are fed into the dynamic CS algorithm for 3D reconstruction. b) Plot of root mean square error (RMSE) as a function of iteration number and time elapsed. As the amount of data increases, the RMSE decreases. c) 2D slices of the 3D reconstruction at various time stamps ($t=1,2$ and completion).}
    \label{fig:overview}
\end{figure} 

\section{Dynamic Compressed Sensing}

Dynamic reconstruction during data collection allows researchers early insight into 3D structure throughout a tomography experiment. Figure \ref{fig:overview} highlights the overall framework for the dynamic CS algorithm. Instead of starting at the end of an experiment, the reconstruction task begins immediately when the first projection is available. As more projections are experimentally acquired, the new information is accommodated as additional constraints in the optimization process (Fig. \ref{fig:overview}a) and improves the reconstructed tomogram quality (Fig. \ref{fig:overview}b). Because the reconstruction process is continuously running throughout the entire data acquisition, which typically take several hours, dynamic compressed sensing is able to produce a high-quality reconstruction before or upon arrival of the final projection (Fig. \ref{fig:overview}c).

For quantitative assessment, a 3D synthetic tomogram ($512^3$ voxels) derived from an experimental gold decorated strontium titanate (Au/SrTiO$_3$) \cite{elliot2017austo} nanoparticle was used to simulate projections along a single axis of rotation. A tilt range of $\pm 75 \degree$ with a two-degree tilt increment ($\Delta \theta = +2 \degree$) approximates the angular range commonly collected in an electron microscope. Samples cannot normally tilt to high angles due to beam shadowing caused by the specimen or holder geometry \cite{banjak2018sirttv}. Poisson noise was added to the projections to give a signal to noise ratio (SNR) of 100. New projections were added every 3 minutes during the reconstruction to mimic typical experimental acquisition rates.

Dynamic compressed sensing produces a 3D tomogram with increasing quality throughout the experiment and by the end is nearly identical to the true SrTiO$_3$ specimen. We can verify the performance by comparing the reconstruction to the ground truth (Fig. \ref{fig:overview}a vs c) and measuring the tomogram's root mean square error (RMSE) (Fig. \ref{fig:overview}b). The reconstruction error reduces with increasing number of iterations and projections acquired. The RMSE curve has a non-smooth, staircase structure when new projections are incorporated into the data vector. Early in the data acquisition process ($< 1$ hr, 14 projections), the limited tilt range results in expected streaking and blurring of the tomogram. As more projections are gathered, additional features are imposed on the 3D volume. Internal voids within cubic SrTiO$_3$ nanoparticles become visible and small Au nanoparticles on the surface become less elongated. Towards the end of an experiment ($> 3$ hrs, 45 projections), the SrTiO$_3$ reconstruction qualitatively matches the true object.

\section{Evaluating Convergence}
\label{sec:convergence}
As dynamic CS progresses, in both iterations and the number of projections, it reduces RMSE but eventually diverges from the optimal solution with minimal error and approaches a solution defined by the optimization problem (Eq. \ref{eq:1}). Iterative algorithms (e.g. Kaczmarz, Landweber, or Cimmino Method) typically deviate from solutions with minimum RMSE when applied to noisy data \cite{hansen2014semiconv}. While RMSE is a useful quantitative measure to assess reconstruction performance, it does not match the visually desired solution \cite{yi2017cslimits}. Moreover, computing RMSE requires knowledge of the true 3D specimen structure. 

\begin{figure}[ht]
    \includegraphics[width=\columnwidth]{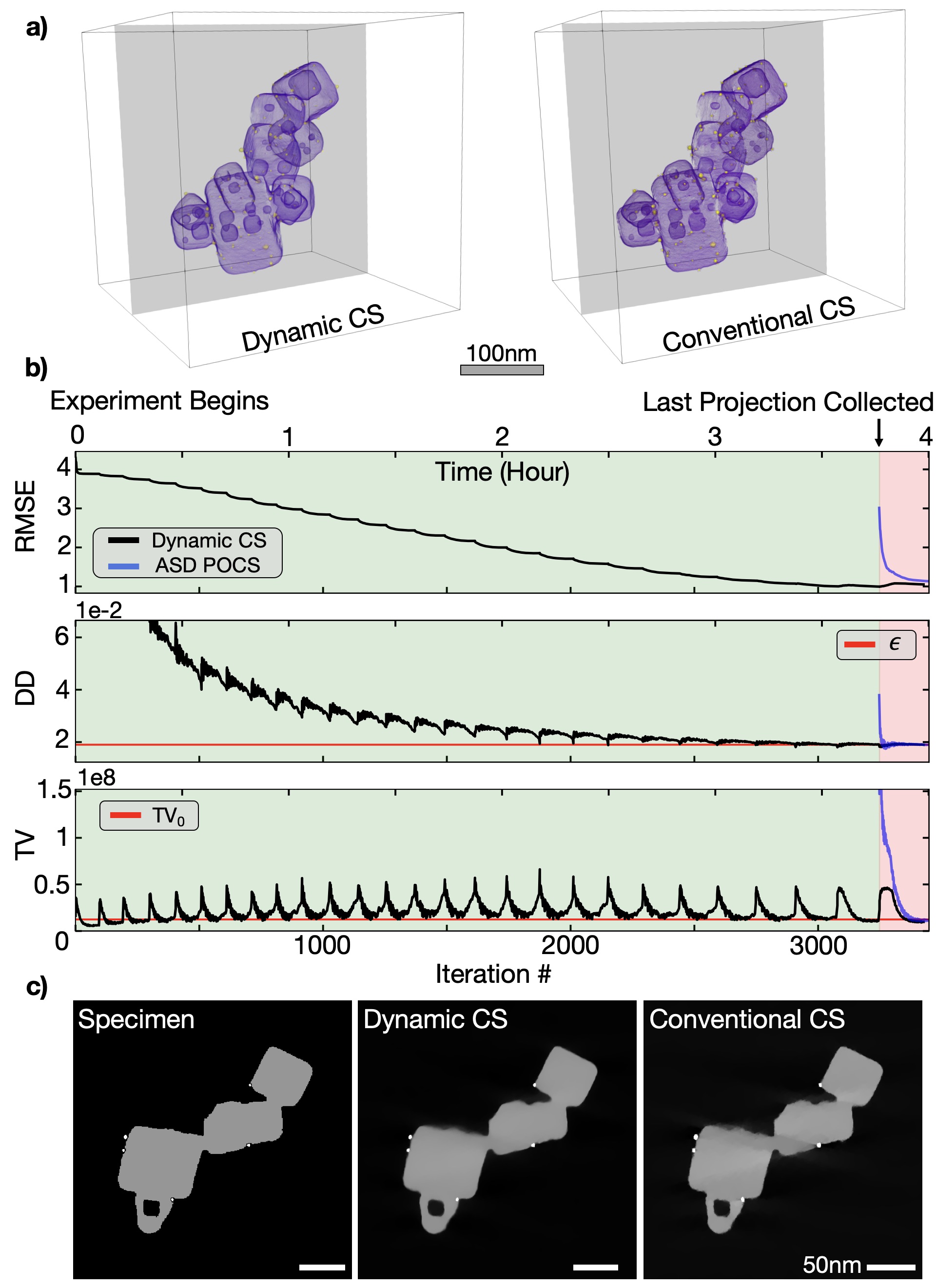}
     \caption{\textbf{Dynamic CS convergence.} a) 3D volume rendering with yellow indicating high intensity (Au) and purple representing low intensity (SrTiO$_3$). These tomograms were constructed under the dynamic (left) and traditional (right) CS framework. b) Plots of data distance (DD), total variation (TV) and RMSE for both the dynamic algorithm (black) and conventional (blue). c) 2D cross-sections of the final output for each 3D reconstruction and the test object.}
    \label{fig:simulation_convergence}
\end{figure}

In real experiments the true solution is unknown and a tomogram's RMSE cannot be measured. Instead, the progression of a reconstruction's TV and data distance (DD$=\|A \hat{\tbf{x}}-b\|_2$) can be utilized to assess the convergence towards an optimal solution.  Figure \ref{fig:simulation_convergence} plots RMSE, DD and TV vs. time for the Au-SrTiO$_3$ phantom nanoparticle during a dynamic compressed sensing reconstruction. Throughout an experiment (shaded green), the data distance trends downward to the specified data tolerance, $\epsilon$ (red line) indicating stable convergence. The incorporation of new data creates sharp discontinuities in DD and TV. Unlike RMSE which drops with the addition of new projections, DD and TV momentarily rise sharply because ASD-POCS is attempting to minimize the distance between DD and $\epsilon$ by iteratively adjusting the weights between data fidelity and regularization. After the arrival of new data, the algorithm will sufficiently converge to a solution within $\sim$125 iterations. Dynamic CS performs best when there are enough iterations to satisfy its data tolerance constraint (DD $\simeq \epsilon$) before new projections are introduced.  If additional projections are added too quickly, the overall convergence may drift (See Supplemental Fig. \ref{fig:full_convergence}) and the algorithm will be unable to reach its optimal solution by experimental completion.


 Dynamic CS reliably produces high reconstruction quality with stable convergence---even before the experiment is complete (e.g. 80\%, 3.5 hrs). After the last projection is collected (shaded red) the RMSE, DD, and TV for dynamic compressed sensing converges $\sim$50\% faster than traditional compressed sensing (blue curve) that starts reconstruction after all data is collected. Dynamic CS benefits from a significantly closer optimal solution that allows for faster convergence when new data arrives. The final reconstruction produced by dynamic CS is indistinguishable from traditional CS and the solution converges to the true object's total variation (TV$_0$) and $\epsilon$.

\section{Dynamic CS for Experimental Electron Tomography}
\begin{figure}[ht]
    \includegraphics[width=\columnwidth]{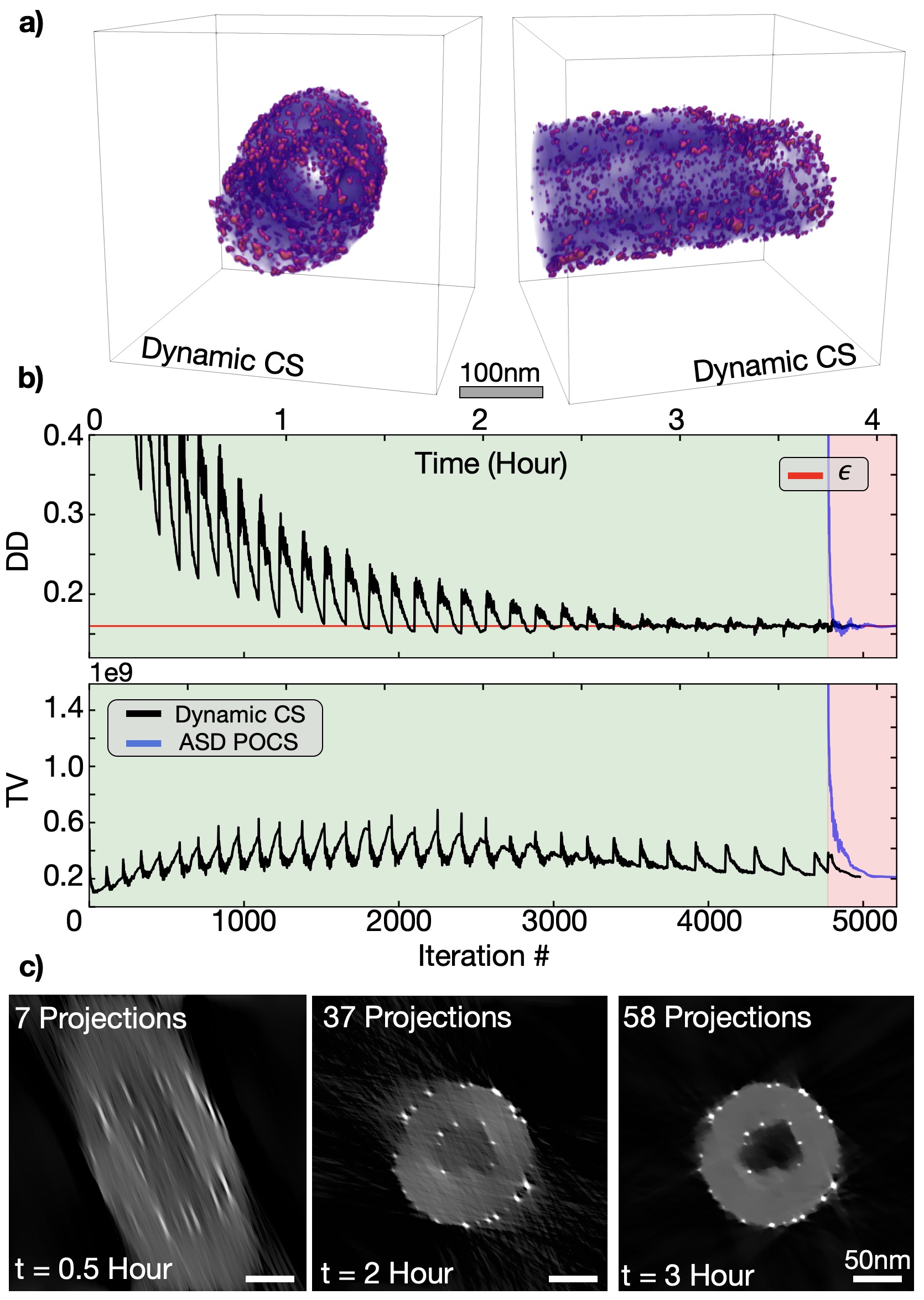}
    \caption{\textbf{Experimental demonstration of dynamic CS.} a) 3D volume rendering of Pt nanoparticles on a carbon nanofiber reconstructed under the dynamic CS framework. Purple indicates low intensity nanofiber and orange-red indicates high intensity Pt nanoparticles. b) Plots of DD and TV vs. time and iterations. c) 2D cross-sections reconstruction at various time stamps ($t=0.5, 2$ and $3$ hours). We can clearly visualize the hollow center and circular Pt nanoparticles decorated on the surface an almost an hour prior to completion.}
    \label{fig:experimental_convergence}
\end{figure}

Dynamic CS can accurately reconstruct experimental electron tomography data. Figure \ref{fig:experimental_convergence} shows dynamic CS reconstruction of platinum (Pt) nanoparticles decorated on a carbon nanofiber. The experimental data had a limited a tilt range of $\pm75\degree$ and a two-degree increment ($\Delta \theta = 2 $). The data, first acquired by Padgett et al.~\cite{elliot2017austo, levin2016etdata}, was introduced into the reconstruction every 3 minutes to simulate experimental acquisition rates. The true object is unknown for experimental data. However, the convergence of DD and TV exhibit similar behaviors to simulation studies (Fig \ref{fig:simulation_convergence}b). That is, DD is progressively driven down to $\epsilon$ and TV is iteratively converging to a minima.

Internal cavities in the carbon nanofiber and the platinum nanoparticles become visible within half the time to complete the experiment (e.g. 2 hrs). Details of the full physical structure becomes clearly visible within 62\% of the complete acquisition (47 projections). The small Pt-nanoparticles are well resolved and show minimum elongation due to the missing wedge. The full structure is clearly visible about an hour prior to completion. 3D visualizations of the final reconstructions are highlighted in Fig. \ref{fig:experimental_convergence}a. 

\section{Dynamic Parameter Tuning}
Selecting $\epsilon$ often requires computing several reconstructions and ultimately relies on the scientist's judgement. Here we show dynamic compressed sensing allows $\epsilon$ to be tightened (decreased) or loosened (increased) mid-reconstruction. Furthermore, the data-constraint can be reversibly adjusted---reflecting stable and convex convergence. Generally speaking, selection of the data consistency constraint, $\epsilon$, depends on the SNR as it accommodates all sources of data inconsistency (e.g. noise) and ensures re-projections are within a given $\ell_2$ distance from the actual (experimental) data \cite{zhang2011ctopt}. Dynamic parameter tuning allows researchers to more efficiently dial in the optimal parameter value. If  $\epsilon$ is too low the reconstruction appears noisy; if $\epsilon$ is too high, detail and resolution is degraded.

\begin{figure*}[ht]
    \includegraphics[width=0.76\linewidth]{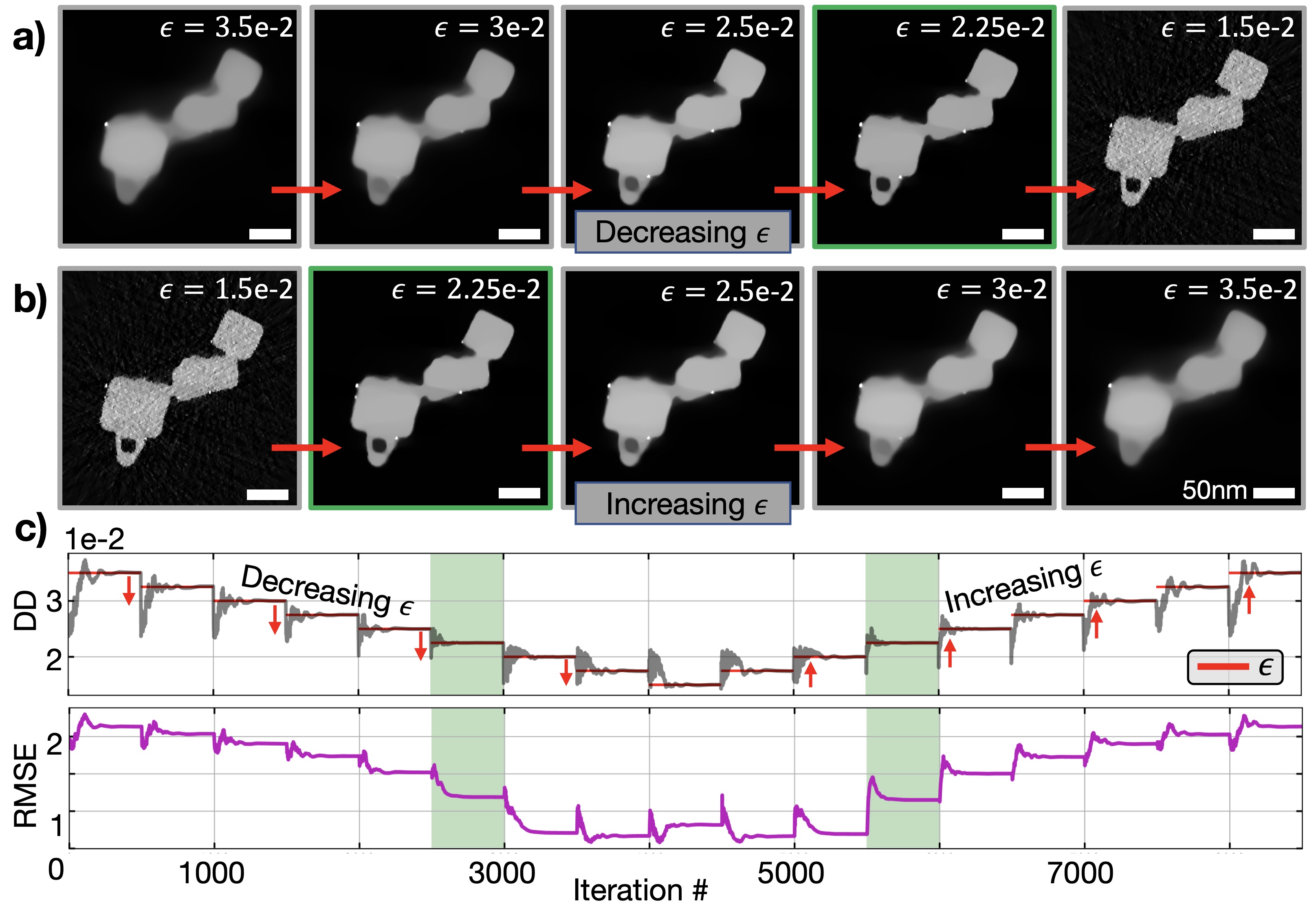}
    \caption{\textbf{Dynamic manipulation of data tolerance parameter.} a) Tightening (reducing) $\epsilon$ decreases the regularization weight which gradually produces sharper 3D tomograms. b) Loosening (increasing) $\epsilon$ allows for there to be more regularization and produce smoother 3D tomograms. c) Plots demonstrating the progression of DD (gray) and RMSE (magenta) vs iteration. As the calculation progresses, scientists can manipulate regularization without having to reset the algorithm. The visually desirable solution is obtained when $\epsilon=0.0225$, highlighted in green. Reducing $\epsilon$ below this value produces noisy reconstructions. Overestimating $\epsilon$ blurs away fine features such as the Pt nanoparticles and internal voids.}
    \label{fig:dynamic_epsilon}
\end{figure*}

Figure \ref{fig:dynamic_epsilon} demonstrates that manipulating $\epsilon$ throughout the reconstruction consistently converges in stable results. Here, we reconstructed a fully sampled Au/SrTiO$_3$ tilt series and waited until all of the optimization parameters (RMSE, DD, and TV) were fully converged prior to perturbing the data constraint. During iterations: 0--4,500 we reduced $\epsilon$ eight times by $-0.025$ before increasing by +0.025 eight times during iterations 4,500--8,500. For this dataset, we found small perturbations ($|\Delta \epsilon| \leq 0.05$) guarantee convergence within $\sim$100 iterations. Dynamic CS reliably converges to solutions defined by the final $\epsilon$ chosen. Whether incrementally increased or decreased, the final value of $\epsilon$ determines a nearly unique solution, which can be seen both visually (Fig \ref{fig:dynamic_epsilon}a,b) and quantitatively in the RMSE \ref{fig:dynamic_epsilon}c. Note that the minimal RMSE ($\epsilon \simeq 0.0175$) retains grainy artifacts and does not produce a desirable reconstruction due to Au particle's high intensities. As discussed by Jiang, \textit{et. al.}, the visually appealing result is generally obtained from a slightly larger $\epsilon$ \cite{yi2017cslimits}. We observed similar data-tolerance properties, the visually desirable reconstruction (highlighted in green Fig. \ref{fig:dynamic_epsilon}) occurs at $\epsilon=0.025$. 


\section{Parallelization \& Performance}
Computational efficiency is key to the success of dynamic compressed sensing. As discussed previously (Section \ref{sec:convergence}), the reconstruction process should reach stable solutions before incorporating new projections. We emphasize that the overall computation should always be faster than the data acquisition, so that by performing the image reconstruction on the fly, we can obtain the reconstructed image almost right after the experiment finishes. This is seen in all the previous cases we have presented, in which RMSE converges to a plateau value before the arrival of a new projection. However, as the size of the object increases, the computational complexity grows as $O(N^3)$, where $N$ is the size of the object in each dimension; the experimental time, however, grows only as $O(N^2)$. Therefore, for large physical systems ($>2048^3$), single laptop/desktop or workstation is not powerful enough for dynamic compressed sensing. To overcome this problem, we deployed high performance computing (HPC) resources at Theta, a Cray XC40 11.69 petaflops supercomputer at Argonne Leadership Computing Facility.

We use Massage Passing Interface (MPI) \cite{Forum93mpi:a} to parallelize our code across different nodes. In particular, we distribute the 2D slices to different MPI processes. There are two dominant computations involved in our algorithm: (1) ART – for  minimizing  $\| A\hat{\tbf{x}} - b \|_2$; (2) TV – for calculating the TV gradient of the object, $\|\hat{\tbf{x}}\|_\text{TV}$. For ART, the computation is independent for different slices and no data exchange is needed among the processes. For TV, each process only needs to send the first and last slices owned by that process to the two nearby processes respectively. The communication overhead thus is minimal. We expect our algorithm to scale efficiently in HPC supercomputers. Besides using MPI for the inter-node parallelism, for the intranode parallelism, we use OpenMP \cite{OpenMP} to parallelize the computation intensive loops.

\section{Discussion \& Conclusion}

Across all varieties of compressed sensing tomography, the optimization process begins after all data has been collected. In this study we demonstrated these methods are compatible with a dynamic, continuously running reconstruction framework which automatically accommodates newly measured data. The algorithm allows for real-time analysis of 3D specimens as an experiment progresses and produces a high-quality reconstruction upon completion of data collection. Although conventional reconstruction algorithms, such as weighted back projection already provide real-time reconstructions due to their simplicity, the quality is significantly inferior to compressed sensing for electron tomography---especially whenever there is limited to moderate number of projections. 

Dynamic CS leverages the time it takes to experimentally acquire projection data in electron tomography to compute high-quality 3D specimen structure. Evaluating dynamic CS against a fully-sampled conventional CS reconstruction, tomogram error is reduced by 27\% in a Au/STiO$_3$ dataset. The successive intermediate updates promote early convergence and it was shown to reconstruct fine features within 62\% of the entire experimental acquisition time. The data tolerance parameter ($\epsilon$) can be interactively tightened or loosened mid-process to mitigate the complexities associated with selecting optimal regularization. Our numerical results illustrate that the proposed dynamic approach practically accelerates the convergence rate of conventional CS algorithms up to a factor of 3. Large-scale reconstructions are achieved with a massively parallelized implementation that benefits from the use of a high-performance computing cluster. The proposed framework builds a foundation for developing more sophisticated algorithms that incorporate projection alignment \cite{odstrcil2019alignment} and machine learning techniques \cite{xin2019deeplearning}. This provides new opportunities for other applications such as hyperspectral or ptychographic tomography---currently active areas of research.  Combined with real-time 3D visualization tools, dynamic CS can enable real-time electron tomography. Scientists can visualize intermediate results with the high-fidelity of compressed sensing as projections are gathered to directly assess detailed specimen structure and optimize experimental parameters.

\begin{acknowledgments}
This work was supported by the Department of Energy (DOE) Office of Science (DE-SC0011385) and Argonne Data Science Program (ADSP). Simulations made use of the Advanced Research Computing Technology Services’ shared high-performance computing at the University of Michigan and Argonne Leadership Computing Facility (ALCF) at the Argonne National Laboratory. ALCF is a DOE Office of Science User Facility supported under Contract DE-AC02-06CH11357. Hardware provided by NVidia GPU Science Grant.
\end{acknowledgments}

\section*{Author contributions statement}

Computational experiments and tomographic reconstruction was conducted by J.S., Y.J., and R.H. Code was optimized and parallelized by H.Z. Data analysis and interpretation was carried out by J.S., H.Z., M.H., Y.J, and R.H. All authors reviewed the manuscript.

\bibliography{main}

\begin{thebibliography}{32}%
\makeatletter
\providecommand \@ifxundefined [1]{%
 \@ifx{#1\undefined}
}%
\providecommand \@ifnum [1]{%
 \ifnum #1\expandafter \@firstoftwo
 \else \expandafter \@secondoftwo
 \fi
}%
\providecommand \@ifx [1]{%
 \ifx #1\expandafter \@firstoftwo
 \else \expandafter \@secondoftwo
 \fi
}%
\providecommand \natexlab [1]{#1}%
\providecommand \enquote  [1]{``#1''}%
\providecommand \bibnamefont  [1]{#1}%
\providecommand \bibfnamefont [1]{#1}%
\providecommand \citenamefont [1]{#1}%
\providecommand \href@noop [0]{\@secondoftwo}%
\providecommand \href [0]{\begingroup \@sanitize@url \@href}%
\providecommand \@href[1]{\@@startlink{#1}\@@href}%
\providecommand \@@href[1]{\endgroup#1\@@endlink}%
\providecommand \@sanitize@url [0]{\catcode `\\12\catcode `\$12\catcode
  `\&12\catcode `\#12\catcode `\^12\catcode `\_12\catcode `\%12\relax}%
\providecommand \@@startlink[1]{}%
\providecommand \@@endlink[0]{}%
\providecommand \url  [0]{\begingroup\@sanitize@url \@url }%
\providecommand \@url [1]{\endgroup\@href {#1}{\urlprefix }}%
\providecommand \urlprefix  [0]{URL }%
\providecommand \Eprint [0]{\href }%
\providecommand \doibase [0]{https://doi.org/}%
\providecommand \selectlanguage [0]{\@gobble}%
\providecommand \bibinfo  [0]{\@secondoftwo}%
\providecommand \bibfield  [0]{\@secondoftwo}%
\providecommand \translation [1]{[#1]}%
\providecommand \BibitemOpen [0]{}%
\providecommand \bibitemStop [0]{}%
\providecommand \bibitemNoStop [0]{.\EOS\space}%
\providecommand \EOS [0]{\spacefactor3000\relax}%
\providecommand \BibitemShut  [1]{\csname bibitem#1\endcsname}%
\let\auto@bib@innerbib\@empty
\bibitem [{\citenamefont {David}\ and\ \citenamefont
  {Klug}(1968)}]{klug1968et}%
  \BibitemOpen
  \bibfield  {author} {\bibinfo {author} {\bibfnamefont {R.}~\bibnamefont
  {David}}\ and\ \bibinfo {author} {\bibfnamefont {A.}~\bibnamefont {Klug}},\
  }\href {https://doi.org/10.1038/217130a0} {\bibfield  {journal} {\bibinfo
  {journal} {Nature}\ }\textbf {\bibinfo {volume} {217}},\ \bibinfo {pages}
  {130} (\bibinfo {year} {1968})}\BibitemShut {NoStop}%
\bibitem [{\citenamefont {Fernandez}\ \emph {et~al.}(2010)\citenamefont
  {Fernandez}, \citenamefont {Zuber}, \citenamefont {Elisabeth}, \citenamefont
  {Cyrklaff}, \citenamefont {Baumeister},\ and\ \citenamefont
  {Lucic}}]{2010cryoem}%
  \BibitemOpen
  \bibfield  {author} {\bibinfo {author} {\bibfnamefont {R.}~\bibnamefont
  {Fernandez}}, \bibinfo {author} {\bibfnamefont {B.}~\bibnamefont {Zuber}},
  \bibinfo {author} {\bibfnamefont {U.}~\bibnamefont {Elisabeth}}, \bibinfo
  {author} {\bibfnamefont {M.}~\bibnamefont {Cyrklaff}}, \bibinfo {author}
  {\bibfnamefont {W.}~\bibnamefont {Baumeister}},\ and\ \bibinfo {author}
  {\bibfnamefont {V.}~\bibnamefont {Lucic}},\ }\href
  {https://doi.org/10.1083/jcb.200908082} {\bibfield  {journal} {\bibinfo
  {journal} {J. Cell Biol.}\ }\textbf {\bibinfo {volume} {188}},\ \bibinfo
  {pages} {145} (\bibinfo {year} {2010})}\BibitemShut {NoStop}%
\bibitem [{\citenamefont {Yu}\ \emph {et~al.}(2011)\citenamefont {Yu},
  \citenamefont {Xin}, \citenamefont {Hovden}, \citenamefont {Wang},
  \citenamefont {Rus}, \citenamefont {Mundy}, \citenamefont {Muller},\ and\
  \citenamefont {Abruña}}]{yu2012fuelcell}%
  \BibitemOpen
  \bibfield  {author} {\bibinfo {author} {\bibfnamefont {Y.}~\bibnamefont
  {Yu}}, \bibinfo {author} {\bibfnamefont {H.}~\bibnamefont {Xin}}, \bibinfo
  {author} {\bibfnamefont {R.}~\bibnamefont {Hovden}}, \bibinfo {author}
  {\bibfnamefont {D.}~\bibnamefont {Wang}}, \bibinfo {author} {\bibfnamefont
  {E.}~\bibnamefont {Rus}}, \bibinfo {author} {\bibfnamefont {J.}~\bibnamefont
  {Mundy}}, \bibinfo {author} {\bibfnamefont {D.}~\bibnamefont {Muller}},\ and\
  \bibinfo {author} {\bibfnamefont {H.}~\bibnamefont {Abruña}},\ }\href
  {https://doi.org/10.1021/nl203920s} {\bibfield  {journal} {\bibinfo
  {journal} {Nano Letters}\ }\textbf {\bibinfo {volume} {12}},\ \bibinfo
  {pages} {4417} (\bibinfo {year} {2011})}\BibitemShut {NoStop}%
\bibitem [{\citenamefont {Sai}\ \emph {et~al.}(2013)\citenamefont {Sai},
  \citenamefont {Tan}, \citenamefont {Hur}, \citenamefont {Asenath-Smith},
  \citenamefont {Hovden}, \citenamefont {Jiang}, \citenamefont {Riccio},
  \citenamefont {Muller}, \citenamefont {Elser}, \citenamefont {Estroff},
  \citenamefont {Gruner},\ and\ \citenamefont {Wiesner}}]{sai2013copolymer}%
  \BibitemOpen
  \bibfield  {author} {\bibinfo {author} {\bibfnamefont {H.}~\bibnamefont
  {Sai}}, \bibinfo {author} {\bibfnamefont {K.~W.}\ \bibnamefont {Tan}},
  \bibinfo {author} {\bibfnamefont {K.}~\bibnamefont {Hur}}, \bibinfo {author}
  {\bibfnamefont {E.}~\bibnamefont {Asenath-Smith}}, \bibinfo {author}
  {\bibfnamefont {R.}~\bibnamefont {Hovden}}, \bibinfo {author} {\bibfnamefont
  {Y.}~\bibnamefont {Jiang}}, \bibinfo {author} {\bibfnamefont
  {M.}~\bibnamefont {Riccio}}, \bibinfo {author} {\bibfnamefont
  {D.}~\bibnamefont {Muller}}, \bibinfo {author} {\bibfnamefont
  {V.}~\bibnamefont {Elser}}, \bibinfo {author} {\bibfnamefont
  {L.}~\bibnamefont {Estroff}}, \bibinfo {author} {\bibfnamefont
  {S.}~\bibnamefont {Gruner}},\ and\ \bibinfo {author} {\bibfnamefont
  {U.}~\bibnamefont {Wiesner}},\ }\href
  {https://doi.org/10.1126/science.1238159} {\bibfield  {journal} {\bibinfo
  {journal} {Science}\ }\textbf {\bibinfo {volume} {341}},\ \bibinfo {pages}
  {530} (\bibinfo {year} {2013})}\BibitemShut {NoStop}%
\bibitem [{\citenamefont {Ercius}\ \emph {et~al.}(2009)\citenamefont {Ercius},
  \citenamefont {Gignac},\ and\ \citenamefont {Muller}}]{ercius2009cu}%
  \BibitemOpen
  \bibfield  {author} {\bibinfo {author} {\bibfnamefont {P.}~\bibnamefont
  {Ercius}}, \bibinfo {author} {\bibfnamefont {L.}~\bibnamefont {Gignac}},\
  and\ \bibinfo {author} {\bibfnamefont {D.}~\bibnamefont {Muller}},\ }\href
  {https://doi.org/10.1017/S143192760909028X} {\bibfield  {journal} {\bibinfo
  {journal} {Microscopy and Microanalysis}\ }\textbf {\bibinfo {volume} {15}},\
  \bibinfo {pages} {244} (\bibinfo {year} {2009})}\BibitemShut {NoStop}%
\bibitem [{\citenamefont {Jiang}\ \emph {et~al.}(2017)\citenamefont {Jiang},
  \citenamefont {Padgett}, \citenamefont {Hovden},\ and\ \citenamefont
  {Muller}}]{yi2017cslimits}%
  \BibitemOpen
  \bibfield  {author} {\bibinfo {author} {\bibfnamefont {Y.}~\bibnamefont
  {Jiang}}, \bibinfo {author} {\bibfnamefont {E.}~\bibnamefont {Padgett}},
  \bibinfo {author} {\bibfnamefont {R.}~\bibnamefont {Hovden}},\ and\ \bibinfo
  {author} {\bibfnamefont {D.}~\bibnamefont {Muller}},\ }\href
  {https://doi.org/10.1016/j.ultramic.2017.12.010} {\bibfield  {journal}
  {\bibinfo  {journal} {Ultramicroscopy}\ }\textbf {\bibinfo {volume} {186}},\
  \bibinfo {pages} {94} (\bibinfo {year} {2017})}\BibitemShut {NoStop}%
\bibitem [{\citenamefont {Donoho}(2006)}]{donoho2006cs}%
  \BibitemOpen
  \bibfield  {author} {\bibinfo {author} {\bibfnamefont {D.}~\bibnamefont
  {Donoho}},\ }\href {https://doi.org/10.1109/TIT.2006.871582} {\bibfield
  {journal} {\bibinfo  {journal} {IEEE Trans. Inf. Thoer.}\ }\textbf {\bibinfo
  {volume} {52}},\ \bibinfo {pages} {1289} (\bibinfo {year}
  {2006})}\BibitemShut {NoStop}%
\bibitem [{\citenamefont {Candés}\ \emph {et~al.}(2006)\citenamefont
  {Candés}, \citenamefont {Romber},\ and\ \citenamefont {Tao}}]{candes2006cs}%
  \BibitemOpen
  \bibfield  {author} {\bibinfo {author} {\bibfnamefont {E.}~\bibnamefont
  {Candés}}, \bibinfo {author} {\bibfnamefont {J.}~\bibnamefont {Romber}},\
  and\ \bibinfo {author} {\bibfnamefont {T.}~\bibnamefont {Tao}},\ }\href
  {https://doi.org/10.1109/TIT.2005.862083} {\bibfield  {journal} {\bibinfo
  {journal} {IEEE Trans. Inf. Thoer.}\ }\textbf {\bibinfo {volume} {52}},\
  \bibinfo {pages} {489} (\bibinfo {year} {2006})}\BibitemShut {NoStop}%
\bibitem [{\citenamefont {Saghi}\ \emph {et~al.}(2011)\citenamefont {Saghi},
  \citenamefont {Holland}, \citenamefont {Leary}, \citenamefont {Falqui},
  \citenamefont {Bertoni}, \citenamefont {Sederman}, \citenamefont {Gladden},\
  and\ \citenamefont {Midgley}}]{midgley2011csnp}%
  \BibitemOpen
  \bibfield  {author} {\bibinfo {author} {\bibfnamefont {Z.}~\bibnamefont
  {Saghi}}, \bibinfo {author} {\bibfnamefont {D.}~\bibnamefont {Holland}},
  \bibinfo {author} {\bibfnamefont {R.}~\bibnamefont {Leary}}, \bibinfo
  {author} {\bibfnamefont {A.}~\bibnamefont {Falqui}}, \bibinfo {author}
  {\bibfnamefont {G.}~\bibnamefont {Bertoni}}, \bibinfo {author} {\bibfnamefont
  {A.}~\bibnamefont {Sederman}}, \bibinfo {author} {\bibfnamefont
  {L.}~\bibnamefont {Gladden}},\ and\ \bibinfo {author} {\bibfnamefont
  {P.}~\bibnamefont {Midgley}},\ }\href {https://doi.org/10.1021/nl202253a}
  {\bibfield  {journal} {\bibinfo  {journal} {Nano Lett.}\ }\textbf {\bibinfo
  {volume} {11}},\ \bibinfo {pages} {4666} (\bibinfo {year}
  {2011})}\BibitemShut {NoStop}%
\bibitem [{\citenamefont {Bicer}\ \emph {et~al.}(2015)\citenamefont {Bicer},
  \citenamefont {Gursoy}, \citenamefont {Kettimuthu}, \citenamefont {De~Carlo},
  \citenamefont {Agarwal},\ and\ \citenamefont {Foster}}]{bicer2015alcftomo}%
  \BibitemOpen
  \bibfield  {author} {\bibinfo {author} {\bibfnamefont {T.}~\bibnamefont
  {Bicer}}, \bibinfo {author} {\bibfnamefont {D.}~\bibnamefont {Gursoy}},
  \bibinfo {author} {\bibfnamefont {R.}~\bibnamefont {Kettimuthu}}, \bibinfo
  {author} {\bibfnamefont {F.}~\bibnamefont {De~Carlo}}, \bibinfo {author}
  {\bibfnamefont {G.}~\bibnamefont {Agarwal}},\ and\ \bibinfo {author}
  {\bibfnamefont {I.}~\bibnamefont {Foster}},\ }\href@noop {} {\bibfield
  {journal} {\bibinfo  {journal} {Euro-Par 2015: Parallel Processing}\ }\textbf
  {\bibinfo {volume} {9233}},\ \bibinfo {pages} {289} (\bibinfo {year}
  {2015})}\BibitemShut {NoStop}%
\bibitem [{\citenamefont {Midgley}\ and\ \citenamefont
  {Weyland}(2003)}]{midgley2003ztomo}%
  \BibitemOpen
  \bibfield  {author} {\bibinfo {author} {\bibfnamefont {P.}~\bibnamefont
  {Midgley}}\ and\ \bibinfo {author} {\bibfnamefont {M.}~\bibnamefont
  {Weyland}},\ }\href {https://doi.org/10.1016/S0304-3991(03)00105-0}
  {\bibfield  {journal} {\bibinfo  {journal} {Ultramicroscopy}\ }\textbf
  {\bibinfo {volume} {96}},\ \bibinfo {pages} {413} (\bibinfo {year}
  {2003})}\BibitemShut {NoStop}%
\bibitem [{\citenamefont {Sidky}\ \emph {et~al.}(2011)\citenamefont {Sidky},
  \citenamefont {Duchin},\ and\ \citenamefont {Pan}}]{sidky2011constrainedtv}%
  \BibitemOpen
  \bibfield  {author} {\bibinfo {author} {\bibfnamefont {E.}~\bibnamefont
  {Sidky}}, \bibinfo {author} {\bibfnamefont {Y.}~\bibnamefont {Duchin}},\ and\
  \bibinfo {author} {\bibfnamefont {X.}~\bibnamefont {Pan}},\ }\href
  {https://doi.org/10.1118/1.3560887} {\bibfield  {journal} {\bibinfo
  {journal} {Med. Phys.}\ }\textbf {\bibinfo {volume} {38}},\ \bibinfo {pages}
  {S117} (\bibinfo {year} {2011})}\BibitemShut {NoStop}%
\bibitem [{\citenamefont {Rudin}\ \emph {et~al.}(1992)\citenamefont {Rudin},
  \citenamefont {Osher},\ and\ \citenamefont {Fatemi}}]{osher1992tv}%
  \BibitemOpen
  \bibfield  {author} {\bibinfo {author} {\bibfnamefont {L.}~\bibnamefont
  {Rudin}}, \bibinfo {author} {\bibfnamefont {S.}~\bibnamefont {Osher}},\ and\
  \bibinfo {author} {\bibfnamefont {E.}~\bibnamefont {Fatemi}},\ }\href
  {https://doi.org/10.1016/0167-2789(92)90242-F} {\bibfield  {journal}
  {\bibinfo  {journal} {Physica D: Nonlinear Phenomena}\ }\textbf {\bibinfo
  {volume} {60}},\ \bibinfo {pages} {259} (\bibinfo {year} {1992})}\BibitemShut
  {NoStop}%
\bibitem [{\citenamefont {Sidky}\ \emph {et~al.}(2006)\citenamefont {Sidky},
  \citenamefont {Chien-Min},\ and\ \citenamefont
  {Xiaochuan}}]{sidky2006tomotv}%
  \BibitemOpen
  \bibfield  {author} {\bibinfo {author} {\bibfnamefont {E.}~\bibnamefont
  {Sidky}}, \bibinfo {author} {\bibfnamefont {K.}~\bibnamefont {Chien-Min}},\
  and\ \bibinfo {author} {\bibfnamefont {P.}~\bibnamefont {Xiaochuan}},\
  }\href@noop {} {\bibfield  {journal} {\bibinfo  {journal} {J.X-Ray Sci.
  Technol.}\ }\textbf {\bibinfo {volume} {14}},\ \bibinfo {pages} {119}
  (\bibinfo {year} {2006})}\BibitemShut {NoStop}%
\bibitem [{\citenamefont {Leary}\ \emph {et~al.}(2013)\citenamefont {Leary},
  \citenamefont {Saghi}, \citenamefont {Midgley},\ and\ \citenamefont
  {Holland}}]{midgley2013cs}%
  \BibitemOpen
  \bibfield  {author} {\bibinfo {author} {\bibfnamefont {R.}~\bibnamefont
  {Leary}}, \bibinfo {author} {\bibfnamefont {Z.}~\bibnamefont {Saghi}},
  \bibinfo {author} {\bibfnamefont {P.}~\bibnamefont {Midgley}},\ and\ \bibinfo
  {author} {\bibfnamefont {D.}~\bibnamefont {Holland}},\ }\href
  {https://doi.org/10.1016/j.ultramic.2013.03.019} {\bibfield  {journal}
  {\bibinfo  {journal} {Ultramicroscopy}\ }\textbf {\bibinfo {volume} {131}},\
  \bibinfo {pages} {70} (\bibinfo {year} {2013})}\BibitemShut {NoStop}%
\bibitem [{\citenamefont {Guay}\ \emph {et~al.}(2016)\citenamefont {Guay},
  \citenamefont {Czaja}, \citenamefont {Aronova},\ and\ \citenamefont
  {Leapman}}]{guay2016csbio}%
  \BibitemOpen
  \bibfield  {author} {\bibinfo {author} {\bibfnamefont {M.}~\bibnamefont
  {Guay}}, \bibinfo {author} {\bibfnamefont {W.}~\bibnamefont {Czaja}},
  \bibinfo {author} {\bibfnamefont {M.}~\bibnamefont {Aronova}},\ and\ \bibinfo
  {author} {\bibfnamefont {R.}~\bibnamefont {Leapman}},\ }\href
  {https://doi.org/10.1038/srep27614} {\bibfield  {journal} {\bibinfo
  {journal} {Scientific Reports}\ }\textbf {\bibinfo {volume} {6}},\ \bibinfo
  {pages} {1} (\bibinfo {year} {2016})}\BibitemShut {NoStop}%
\bibitem [{\citenamefont {Sidky}\ and\ \citenamefont
  {Pan}(2008)}]{sidky2008asd}%
  \BibitemOpen
  \bibfield  {author} {\bibinfo {author} {\bibfnamefont {E.}~\bibnamefont
  {Sidky}}\ and\ \bibinfo {author} {\bibfnamefont {X.}~\bibnamefont {Pan}},\
  }\href {https://doi.org/10.1088/0031-9155/53/17/021} {\bibfield  {journal}
  {\bibinfo  {journal} {Pys. Med. Biol.}\ }\textbf {\bibinfo {volume} {53}},\
  \bibinfo {pages} {4777} (\bibinfo {year} {2008})}\BibitemShut {NoStop}%
\bibitem [{\citenamefont {Liu}\ \emph {et~al.}(2017)\citenamefont {Liu},
  \citenamefont {Han},\ and\ \citenamefont {Jin}}]{liu2017alttomo}%
  \BibitemOpen
  \bibfield  {author} {\bibinfo {author} {\bibfnamefont {L.}~\bibnamefont
  {Liu}}, \bibinfo {author} {\bibfnamefont {Y.}~\bibnamefont {Han}},\ and\
  \bibinfo {author} {\bibfnamefont {M.}~\bibnamefont {Jin}},\ }\href
  {https://doi.org/10.1371/journal.pone.0172938} {\bibfield  {journal}
  {\bibinfo  {journal} {PLoS One}\ }\textbf {\bibinfo {volume} {12}},\ \bibinfo
  {pages} {1} (\bibinfo {year} {2017})}\BibitemShut {NoStop}%
\bibitem [{\citenamefont {Padgett}\ \emph {et~al.}(2017)\citenamefont
  {Padgett}, \citenamefont {Hovden}, \citenamefont {DaSilva}, \citenamefont
  {Levin}, \citenamefont {Grazul}, \citenamefont {Hanrath},\ and\ \citenamefont
  {Muller}}]{elliot2017austo}%
  \BibitemOpen
  \bibfield  {author} {\bibinfo {author} {\bibfnamefont {E.}~\bibnamefont
  {Padgett}}, \bibinfo {author} {\bibfnamefont {R.}~\bibnamefont {Hovden}},
  \bibinfo {author} {\bibfnamefont {J.}~\bibnamefont {DaSilva}}, \bibinfo
  {author} {\bibfnamefont {B.}~\bibnamefont {Levin}}, \bibinfo {author}
  {\bibfnamefont {J.}~\bibnamefont {Grazul}}, \bibinfo {author} {\bibfnamefont
  {T.}~\bibnamefont {Hanrath}},\ and\ \bibinfo {author} {\bibfnamefont
  {D.}~\bibnamefont {Muller}},\ }\href
  {https://doi.org/10.1017/S1431927617012764} {\bibfield  {journal} {\bibinfo
  {journal} {Microscopy and Microanalysis}\ }\textbf {\bibinfo {volume} {23}},\
  \bibinfo {pages} {1150} (\bibinfo {year} {2017})}\BibitemShut {NoStop}%
\bibitem [{\citenamefont {Banjak}\ \emph {et~al.}(2018)\citenamefont {Banjak},
  \citenamefont {Grenier}, \citenamefont {Epicier}, \citenamefont {Koneti},
  \citenamefont {Roiban}, \citenamefont {Gay}, \citenamefont {Magnin},
  \citenamefont {Peyrin},\ and\ \citenamefont {Maxim}}]{banjak2018sirttv}%
  \BibitemOpen
  \bibfield  {author} {\bibinfo {author} {\bibfnamefont {H.}~\bibnamefont
  {Banjak}}, \bibinfo {author} {\bibfnamefont {T.}~\bibnamefont {Grenier}},
  \bibinfo {author} {\bibfnamefont {T.}~\bibnamefont {Epicier}}, \bibinfo
  {author} {\bibfnamefont {S.}~\bibnamefont {Koneti}}, \bibinfo {author}
  {\bibfnamefont {L.}~\bibnamefont {Roiban}}, \bibinfo {author} {\bibfnamefont
  {A.}~\bibnamefont {Gay}}, \bibinfo {author} {\bibfnamefont {I.}~\bibnamefont
  {Magnin}}, \bibinfo {author} {\bibfnamefont {F.}~\bibnamefont {Peyrin}},\
  and\ \bibinfo {author} {\bibfnamefont {V.}~\bibnamefont {Maxim}},\ }\href
  {https://doi.org/10.1016/j.ultramic.2018.03.022} {\bibfield  {journal}
  {\bibinfo  {journal} {Ultramicroscopy}\ }\textbf {\bibinfo {volume} {189}},\
  \bibinfo {pages} {109} (\bibinfo {year} {2018})}\BibitemShut {NoStop}%
\bibitem [{\citenamefont {Elfving}\ \emph {et~al.}(2014)\citenamefont
  {Elfving}, \citenamefont {Hansen},\ and\ \citenamefont
  {Nikazad}}]{hansen2014semiconv}%
  \BibitemOpen
  \bibfield  {author} {\bibinfo {author} {\bibfnamefont {T.}~\bibnamefont
  {Elfving}}, \bibinfo {author} {\bibfnamefont {P.~C.}\ \bibnamefont
  {Hansen}},\ and\ \bibinfo {author} {\bibfnamefont {T.}~\bibnamefont
  {Nikazad}},\ }\href {https://doi.org/10.1088/0266-5611/30/5/055007}
  {\bibfield  {journal} {\bibinfo  {journal} {Inverse Problems}\ }\textbf
  {\bibinfo {volume} {30}},\ \bibinfo {pages} {1} (\bibinfo {year}
  {2014})}\BibitemShut {NoStop}%
\bibitem [{\citenamefont {Levin}\ \emph {et~al.}(2016)\citenamefont {Levin},
  \citenamefont {Padgett}, \citenamefont {Chen}, \citenamefont {Scott},
  \citenamefont {Xu}, \citenamefont {Theis}, \citenamefont {Yang},
  \citenamefont {Ophus}, \citenamefont {Zhang}, \citenamefont {Ha},
  \citenamefont {Wang}, \citenamefont {Yu}, \citenamefont {Abruña},
  \citenamefont {Robinson}, \citenamefont {Ercius}, \citenamefont {Kourkoutis},
  \citenamefont {Miao}, \citenamefont {Muller},\ and\ \citenamefont
  {Hovden}}]{levin2016etdata}%
  \BibitemOpen
  \bibfield  {author} {\bibinfo {author} {\bibfnamefont {B.}~\bibnamefont
  {Levin}}, \bibinfo {author} {\bibfnamefont {E.}~\bibnamefont {Padgett}},
  \bibinfo {author} {\bibfnamefont {C.-C.}\ \bibnamefont {Chen}}, \bibinfo
  {author} {\bibfnamefont {M.}~\bibnamefont {Scott}}, \bibinfo {author}
  {\bibfnamefont {R.}~\bibnamefont {Xu}}, \bibinfo {author} {\bibfnamefont
  {W.}~\bibnamefont {Theis}}, \bibinfo {author} {\bibfnamefont
  {Y.}~\bibnamefont {Yang}}, \bibinfo {author} {\bibfnamefont {C.}~\bibnamefont
  {Ophus}}, \bibinfo {author} {\bibfnamefont {H.}~\bibnamefont {Zhang}},
  \bibinfo {author} {\bibfnamefont {D.-H.}\ \bibnamefont {Ha}}, \bibinfo
  {author} {\bibfnamefont {D.}~\bibnamefont {Wang}}, \bibinfo {author}
  {\bibfnamefont {Y.}~\bibnamefont {Yu}}, \bibinfo {author} {\bibfnamefont
  {H.}~\bibnamefont {Abruña}}, \bibinfo {author} {\bibfnamefont
  {R.}~\bibnamefont {Robinson}}, \bibinfo {author} {\bibfnamefont
  {P.}~\bibnamefont {Ercius}}, \bibinfo {author} {\bibfnamefont
  {L.}~\bibnamefont {Kourkoutis}}, \bibinfo {author} {\bibfnamefont
  {J.}~\bibnamefont {Miao}}, \bibinfo {author} {\bibfnamefont {D.}~\bibnamefont
  {Muller}},\ and\ \bibinfo {author} {\bibfnamefont {R.}~\bibnamefont
  {Hovden}},\ }\href {https://doi.org/10.1038/sdata.2016.41} {\bibfield
  {journal} {\bibinfo  {journal} {Scientific Data}\ }\textbf {\bibinfo {volume}
  {3}},\ \bibinfo {pages} {1} (\bibinfo {year} {2016})}\BibitemShut {NoStop}%
\bibitem [{\citenamefont {Zhang}\ \emph {et~al.}(2011)\citenamefont {Zhang},
  \citenamefont {Wang},\ and\ \citenamefont {Xing}}]{zhang2011ctopt}%
  \BibitemOpen
  \bibfield  {author} {\bibinfo {author} {\bibfnamefont {X.}~\bibnamefont
  {Zhang}}, \bibinfo {author} {\bibfnamefont {J.}~\bibnamefont {Wang}},\ and\
  \bibinfo {author} {\bibfnamefont {L.}~\bibnamefont {Xing}},\ }\href
  {https://doi.org/10.1118/1.3533711} {\bibfield  {journal} {\bibinfo
  {journal} {Med. Phys.}\ }\textbf {\bibinfo {volume} {38}},\ \bibinfo {pages}
  {701} (\bibinfo {year} {2011})}\BibitemShut {NoStop}%
\bibitem [{\citenamefont {Clarke}\ \emph {et~al.}(1994)\citenamefont {Clarke},
  \citenamefont {Glendinning},\ and\ \citenamefont {Hempel}}]{Forum93mpi:a}%
  \BibitemOpen
  \bibfield  {author} {\bibinfo {author} {\bibfnamefont {L.}~\bibnamefont
  {Clarke}}, \bibinfo {author} {\bibfnamefont {I.}~\bibnamefont
  {Glendinning}},\ and\ \bibinfo {author} {\bibfnamefont {R.}~\bibnamefont
  {Hempel}},\ }\href@noop {} {\bibfield  {journal} {\bibinfo  {journal}
  {Programming Environments for Massively Parallel Dirstributed Systems}\ ,\
  \bibinfo {pages} {213}} (\bibinfo {year} {1994})}\BibitemShut {NoStop}%
\bibitem [{\citenamefont {Dagum}\ and\ \citenamefont {Menon}(1998)}]{OpenMP}%
  \BibitemOpen
  \bibfield  {author} {\bibinfo {author} {\bibfnamefont {L.}~\bibnamefont
  {Dagum}}\ and\ \bibinfo {author} {\bibfnamefont {R.}~\bibnamefont {Menon}},\
  }\href {https://doi.org/10.1109/99.660313} {\bibfield  {journal} {\bibinfo
  {journal} {IEEE Comput. Sci. Eng.}\ }\textbf {\bibinfo {volume} {5}},\
  \bibinfo {pages} {46–55} (\bibinfo {year} {1998})}\BibitemShut {NoStop}%
\bibitem [{\citenamefont {Odstrcil}\ \emph {et~al.}(2019)\citenamefont
  {Odstrcil}, \citenamefont {Holler}, \citenamefont {Raabe},\ and\
  \citenamefont {Guizar-Sicairos}}]{odstrcil2019alignment}%
  \BibitemOpen
  \bibfield  {author} {\bibinfo {author} {\bibfnamefont {M.}~\bibnamefont
  {Odstrcil}}, \bibinfo {author} {\bibfnamefont {M.}~\bibnamefont {Holler}},
  \bibinfo {author} {\bibfnamefont {J.}~\bibnamefont {Raabe}},\ and\ \bibinfo
  {author} {\bibfnamefont {M.}~\bibnamefont {Guizar-Sicairos}},\ }\href
  {https://doi.org/10.1364/OE.27.036637} {\bibfield  {journal} {\bibinfo
  {journal} {Optics Express}\ }\textbf {\bibinfo {volume} {27}},\ \bibinfo
  {pages} {36637} (\bibinfo {year} {2019})}\BibitemShut {NoStop}%
\bibitem [{\citenamefont {Ding}\ \emph {et~al.}(2019)\citenamefont {Ding},
  \citenamefont {Zhang},\ and\ \citenamefont {Xin}}]{xin2019deeplearning}%
  \BibitemOpen
  \bibfield  {author} {\bibinfo {author} {\bibfnamefont {G.}~\bibnamefont
  {Ding}}, \bibinfo {author} {\bibfnamefont {R.}~\bibnamefont {Zhang}},\ and\
  \bibinfo {author} {\bibfnamefont {H.}~\bibnamefont {Xin}},\ }\href
  {https://doi.org/10.1038/s41598-019-49267-x} {\bibfield  {journal} {\bibinfo
  {journal} {Scientific Reports}\ }\textbf {\bibinfo {volume} {9}},\ \bibinfo
  {pages} {1} (\bibinfo {year} {2019})}\BibitemShut {NoStop}%
\bibitem [{\citenamefont {Gordon}\ \emph {et~al.}(1970)\citenamefont {Gordon},
  \citenamefont {Bender},\ and\ \citenamefont {Herman}}]{gordon1970art}%
  \BibitemOpen
  \bibfield  {author} {\bibinfo {author} {\bibfnamefont {R.}~\bibnamefont
  {Gordon}}, \bibinfo {author} {\bibfnamefont {R.}~\bibnamefont {Bender}},\
  and\ \bibinfo {author} {\bibfnamefont {G.}~\bibnamefont {Herman}},\ }\href
  {https://doi.org/10.1016/0022-5193(70)90109-8} {\bibfield  {journal}
  {\bibinfo  {journal} {J. Theor. Biol.}\ }\textbf {\bibinfo {volume} {29}},\
  \bibinfo {pages} {471} (\bibinfo {year} {1970})}\BibitemShut {NoStop}%
\bibitem [{\citenamefont {Kaczmarz}(1993)}]{kaczmarz1993art}%
  \BibitemOpen
  \bibfield  {author} {\bibinfo {author} {\bibfnamefont {S.}~\bibnamefont
  {Kaczmarz}},\ }\href {https://doi.org/10.1080/00207179308934446} {\bibfield
  {journal} {\bibinfo  {journal} {Int. J. Control}\ }\textbf {\bibinfo {volume}
  {57}},\ \bibinfo {pages} {1269} (\bibinfo {year} {1993})}\BibitemShut
  {NoStop}%
\bibitem [{\citenamefont {Strohmer}\ and\ \citenamefont
  {Vershynin}(2009)}]{strohmer2009rart}%
  \BibitemOpen
  \bibfield  {author} {\bibinfo {author} {\bibfnamefont {T.}~\bibnamefont
  {Strohmer}}\ and\ \bibinfo {author} {\bibfnamefont {R.}~\bibnamefont
  {Vershynin}},\ }\href {https://doi.org/10.1007/s00041-008-9030-4} {\bibfield
  {journal} {\bibinfo  {journal} {J. Fourier Anal. Appl.}\ }\textbf {\bibinfo
  {volume} {15}},\ \bibinfo {pages} {262} (\bibinfo {year} {2009})}\BibitemShut
  {NoStop}%
\bibitem [{\citenamefont {Sørensen}\ and\ \citenamefont
  {Hansen}(2014)}]{hansen2014algebraic}%
  \BibitemOpen
  \bibfield  {author} {\bibinfo {author} {\bibfnamefont {H.}~\bibnamefont
  {Sørensen}}\ and\ \bibinfo {author} {\bibfnamefont {P.~C.}\ \bibnamefont
  {Hansen}},\ }\href {https://doi.org/10.1137/130920642} {\bibfield  {journal}
  {\bibinfo  {journal} {SIAM J. Sci. Comput.}\ }\textbf {\bibinfo {volume}
  {36}},\ \bibinfo {pages} {C524} (\bibinfo {year} {2014})}\BibitemShut
  {NoStop}%
\bibitem [{\citenamefont {Schwartz}\ \emph {et~al.}(2019)\citenamefont
  {Schwartz}, \citenamefont {Jiang}, \citenamefont {Wang}, \citenamefont
  {Aiello}, \citenamefont {Bhattacharya}, \citenamefont {Yuan}, \citenamefont
  {Mi}, \citenamefont {Bassim},\ and\ \citenamefont {Hovden}}]{schwartz2018CS}%
  \BibitemOpen
  \bibfield  {author} {\bibinfo {author} {\bibfnamefont {J.}~\bibnamefont
  {Schwartz}}, \bibinfo {author} {\bibfnamefont {Y.}~\bibnamefont {Jiang}},
  \bibinfo {author} {\bibfnamefont {Y.}~\bibnamefont {Wang}}, \bibinfo {author}
  {\bibfnamefont {A.}~\bibnamefont {Aiello}}, \bibinfo {author} {\bibfnamefont
  {P.}~\bibnamefont {Bhattacharya}}, \bibinfo {author} {\bibfnamefont
  {H.}~\bibnamefont {Yuan}}, \bibinfo {author} {\bibfnamefont {Z.}~\bibnamefont
  {Mi}}, \bibinfo {author} {\bibfnamefont {N.}~\bibnamefont {Bassim}},\ and\
  \bibinfo {author} {\bibfnamefont {R.}~\bibnamefont {Hovden}},\ }\href
  {https://doi.org/10.1017/S1431927619000254} {\bibfield  {journal} {\bibinfo
  {journal} {Microscopy and Microanalysis}\ }\textbf {\bibinfo {volume} {25}},\
  \bibinfo {pages} {705} (\bibinfo {year} {2019})}\BibitemShut {NoStop}%
\end{thebibliography}%

\FloatBarrier
\clearpage
\newpage
\pagebreak

\appendix
\section{Supplemental Figures}
\thispagestyle{plain}
\begin{figure}[ht]
    \includegraphics[width=\columnwidth]{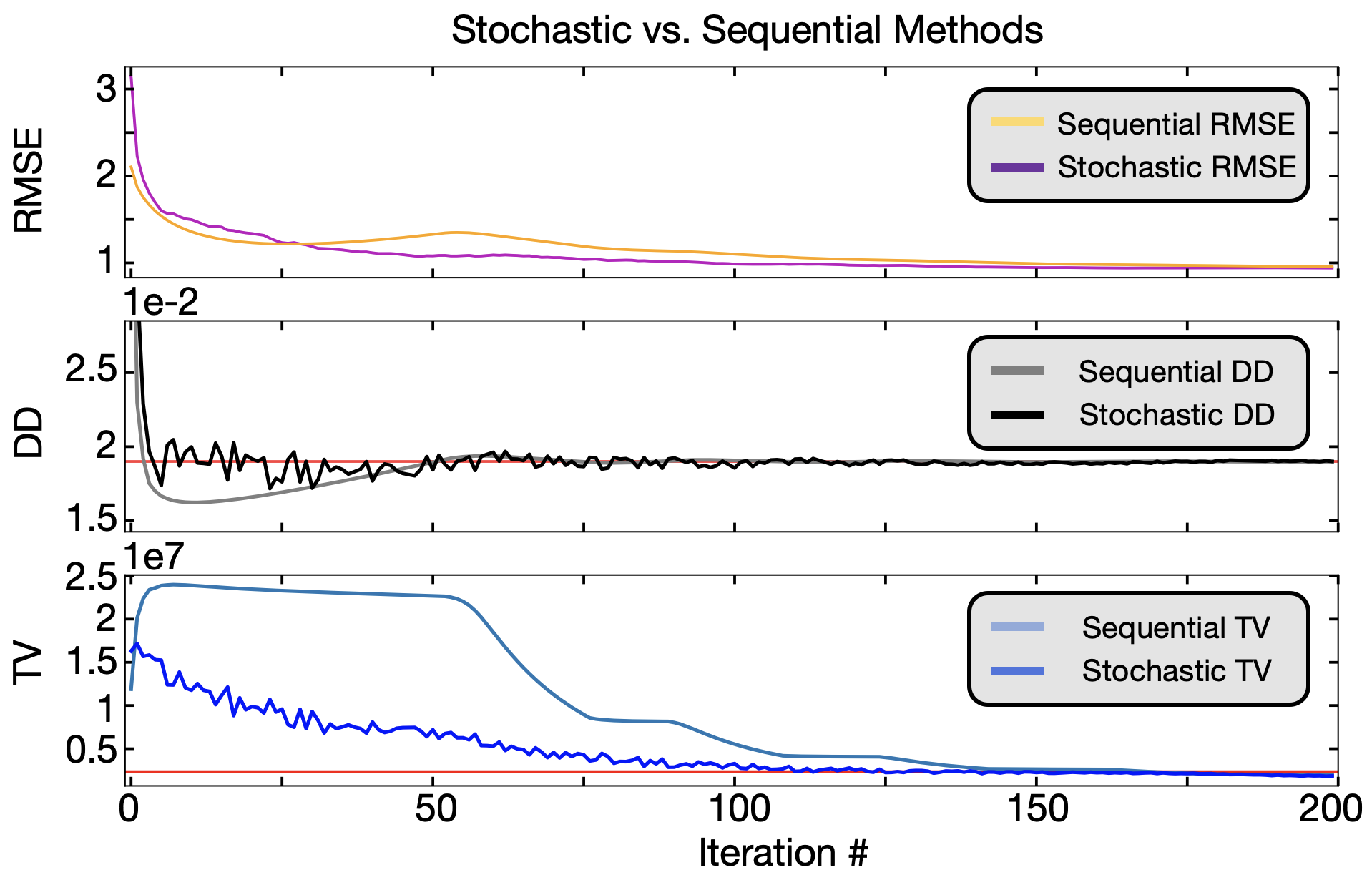}
    \caption{\textbf{Comparison between sequential and randomized ART.} Plots highlighting the convergence of RMSE, DD, and TV for ASD-POCS that use stochastic and sequential ART to enforce data-fidelity. The stochastic variation removes any bias associated with sequentially cycling through the rows of the $A$ which results in faster convergence for both DD and TV.}
    \label{fig:randomized_convergence}
\end{figure} 

\begin{figure}[ht]
    \includegraphics[width=\columnwidth]{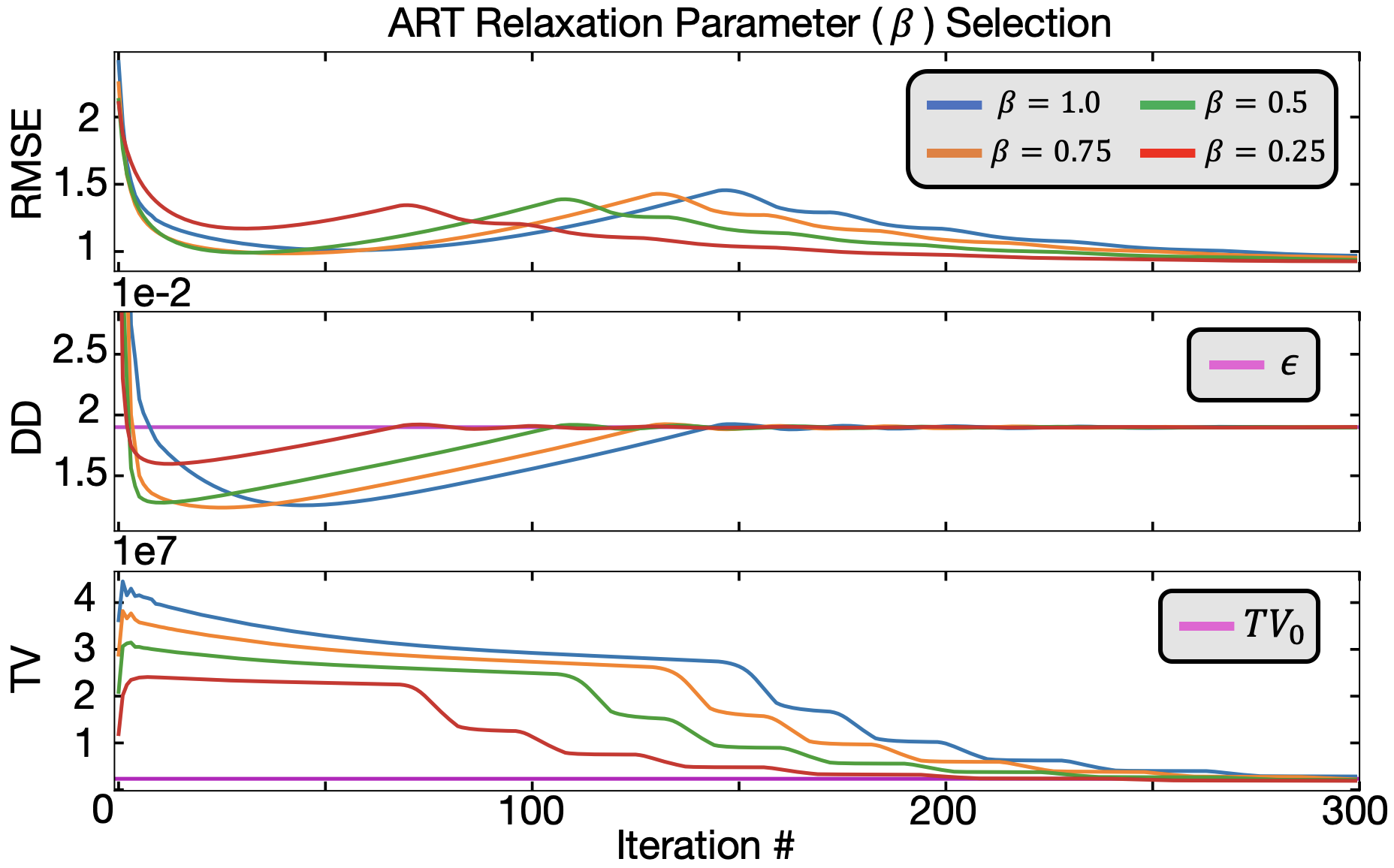}
        \caption{\textbf{ART relaxation parameter selection.} Plots highlighting the convergence of RMSE, DD, and TV for multiple relaxation parameters ($\beta$) in ART update. The choice of $\beta$ has significant impact to the convergence of ASD-POCS reconstruction.}
    \label{fig:beta_convergence}
\end{figure}

\begin{figure}[ht]
    \includegraphics[width=\columnwidth]{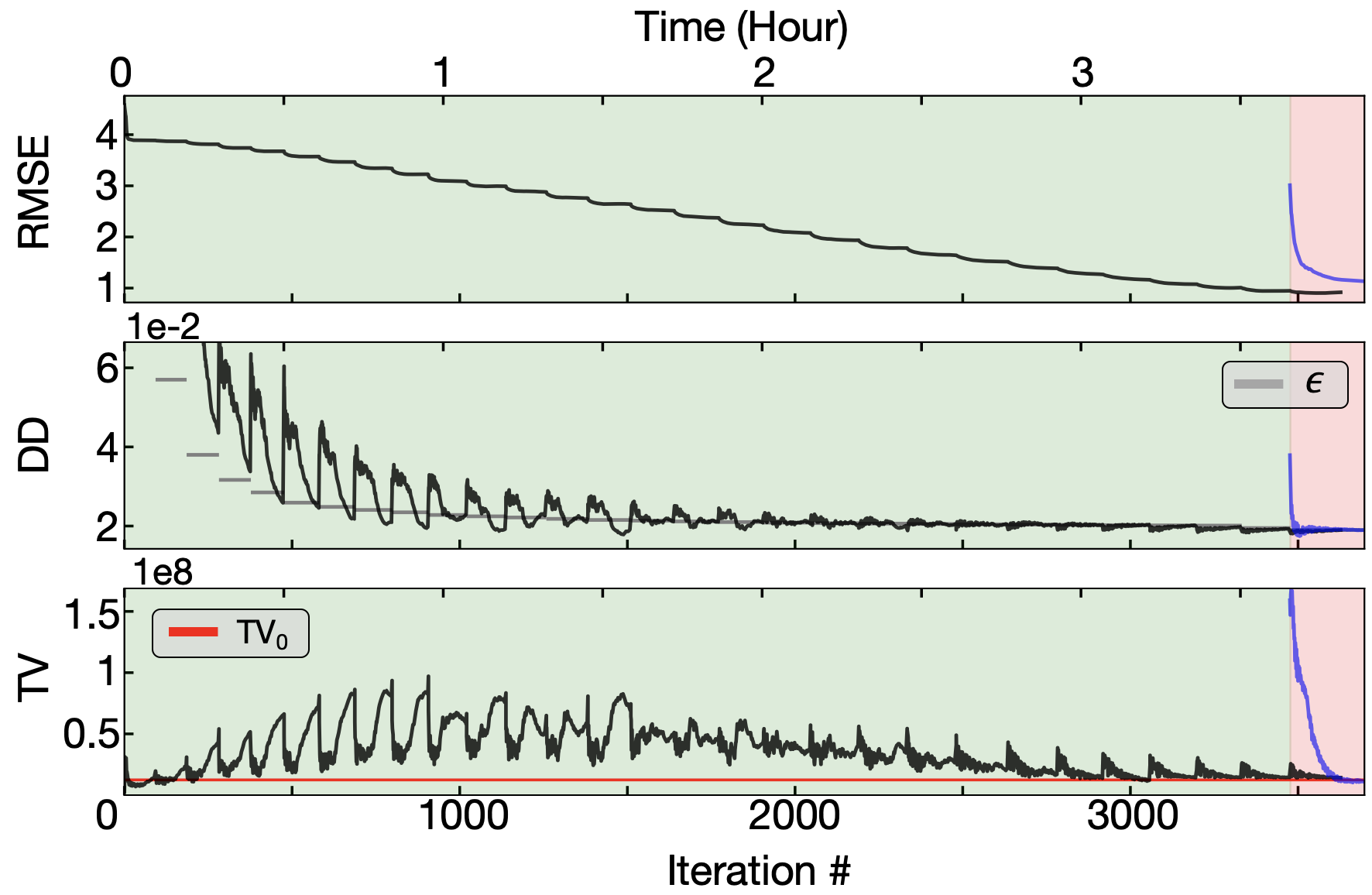}
    \caption{\textbf{CS reconstruction with a dynamic epsilon.} Plots highlighting the convergence of RMSE, DD, and TV for the synthetic Au/SrTiO$_3$ nanoparticle. The reconstruction begins with a loose $\epsilon$ estimate and sequentially tightens the constraint. We see this helps TV converge within 50 iterations.} 
    \label{fig:varying_eps_recon}
\end{figure}

\begin{figure}[ht]
    \includegraphics[width=\columnwidth]{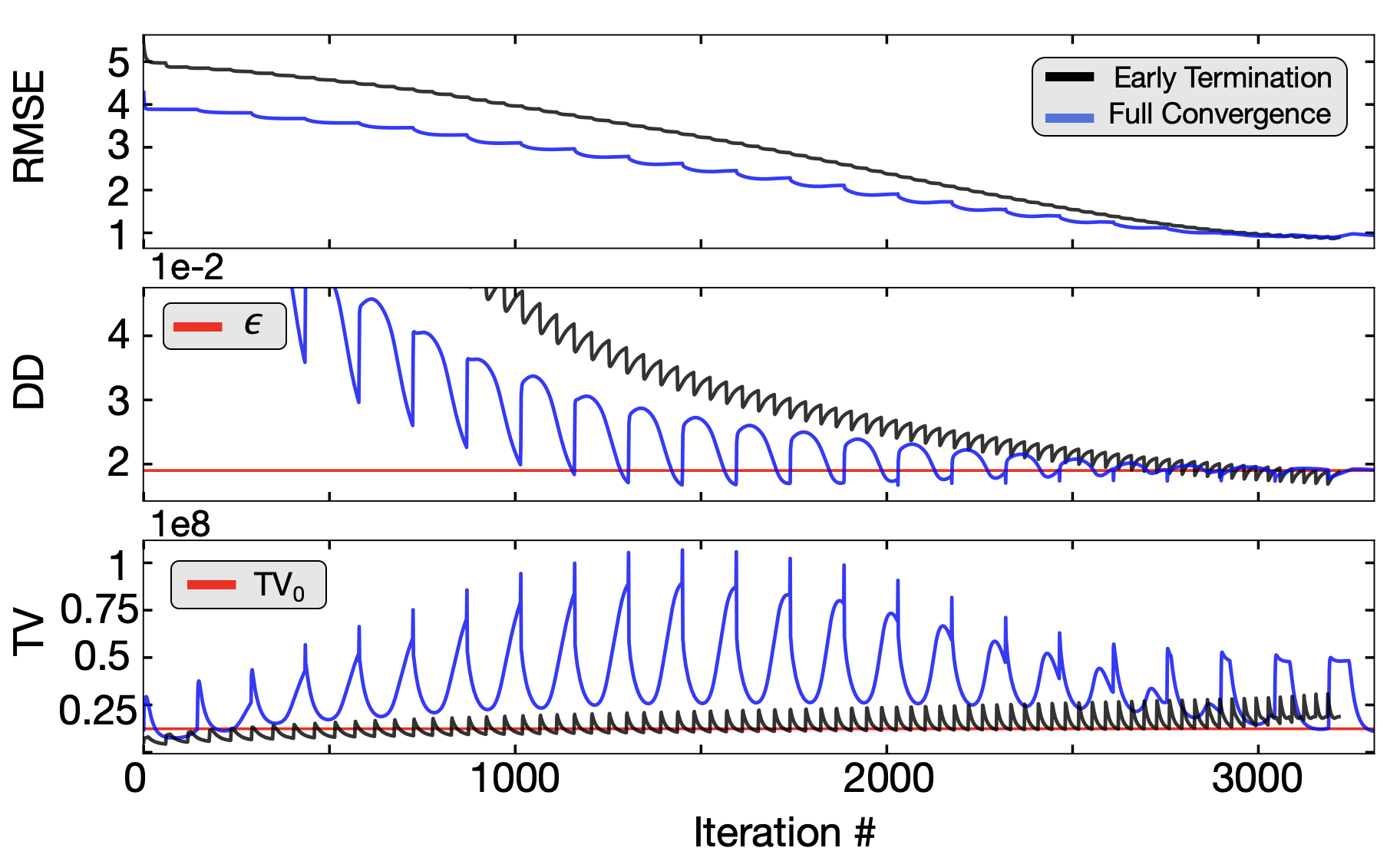}
    \caption{\textbf{Convergence criteria.} Stable (blue) and unstable (black) convergence of RMSE, DD, and TV for the synthetic Au/SrTiO$_3$ nanoparticle. The blue curve shows proper convergence when the reconstruction completes many iterations (125) prior to appending more projections. The black curve shows incomplete convergence when insufficient intermediate iterations are not completed and TV may drift away from the true solution (TV$_0$).}
    \label{fig:full_convergence}
\end{figure} 

\FloatBarrier
\clearpage
\newpage
\pagebreak
\section{Dynamic Compressed Sensing Algorithm Parameters}
\thispagestyle{plain}
The pseudo-code for our proposed algorithm is listed in Supplemental Algorithm \ref{alg:dyn_cs}. Its basic steps are summarized as follows: The outer loop controls the distribution of projections to the data matrix ($b$). For each projection acquisition, the imaging data is appended to $b$ and utilized to further reduce the tomogram error (illustrated in Fig. \ref{fig:overview}a). The 3D reconstruction on lines \ref{alg:asd_start} - \ref{alg:asd_end} is similar to the pseudo-code in \cite{sidky2008asd} with some variation in the relaxation parameter reduction and POCS step. It is essential to allow the 3D reconstruction converge prior to appending more data to $b$. Terminating the construction process will restrict the overall performance and cause convergence parameters to diverge from the true (optimal) solution. (illustrated in Fig. \ref{fig:full_convergence}).

One can enforce data fidelity (POCS) by utilizing Kaczmarz algorithm, also known as the Algebraic Reconstruction Technique (ART). ART is an iterative method that solves inverse problems \cite{gordon1970art,kaczmarz1993art} by sequentially using each row of the measurement matrix ($a_i$) to minimize the tomogram's $\ell_2$ error. Each ART update follows:
\begin{align}
    x^{(k+1)} = x^{(k)} + \beta \frac{b_i - \langle a_i, x^{(k)} \rangle}{\|a_i \|^2} a_i
\end{align} 
where index $k$ is the iteration number and $\beta$ is a relaxation parameter. In this work, instead of sequentially cycling through the rows of $A$,we randomly select rows during ART update to achieve faster convergence \cite{strohmer2009rart} (demonstrated in Fig. \ref{fig:randomized_convergence}). According to literature one can guarantee convergence when $0 < \beta < 2$ \cite{hansen2014algebraic}. For the simulated and experimental objects studied here, we found optimal performance occurs when $\beta < 0.5$. If too large of a relaxation parameter is chosen, the ASD-POCS algorithm requires significantly more iterations to drive the data distance (DD$ =\| A \tbf{x} - \tbf{b}\|_2$) to reach $\epsilon$ (shown in Fig. \ref{fig:beta_convergence}). 

The tomogram's TV has a tendency to expand linearly with increasing projection number. As highlighted in the blue curve in Figure \ref{fig:full_convergence}, the range between the minimal and maximal value grows consistently. To suppress this behavior, we introduced a envelope dampening calculation on line 8. This calculation ensures that when the relaxation parameter is reset, its magnitude decays linearly. Empirically we found that the ratio $\frac{5}{6}$ works consistently across all of our test data sets.

Outside of the influence from $\beta$, there are additional parameters that control the dynamic CS algorithm. Complimentary to $\beta$, there is reduction parameter $\beta_{\text{red}}$ that decays the step-size for ART. Similarly, there is also relaxation and reduction parameter for TV minimization ($\alpha,\alpha_{\text{red}}$). $\alpha_{\text{red}}$ should be less than $\beta_{\text{red}}$ so that the tomogram can maintain the data consistency constraint. The ratio of change in the 3D volume from the ART and TV stage is evaluated against $r_{\text{max}}$ (usually $<1.0$). There also two variants total iteration number: minimum iteration to achieve convergence ($N_{\text{min}}$) and TV gradient descent ($ng$). Previous results show that 150 iterations is sufficient to achieve proper convergence \cite{schwartz2018CS}. Overall these parameters rarely need to tuned too drastically and the reconstruction quality will mostly depend on $\epsilon$ selection.  

\onecolumngrid

\setlength{\intextsep}{3pt}
\section{Pseudo-Code for Dynamic Compressed Sensing Tomography}
\thispagestyle{plain}
\begin{algorithm} [H]
	\caption{Dynamic ASD - POCS}
	\label{alg:dyn_cs}
	\begin{algorithmic}[1]
		\State \tbf{Input:} $b \in \mathbb{R}^3 \rightarrow \text{Projections}$
		\State \tbf{Input:} $\epsilon = 0.2 \rightarrow \text{Data Tolerance}$
		\State $\tbf{Initialize algorithm parameters: } $
		\State $\beta_0 = 0.5, ~ \beta_{\text{red}} = 0.98,  \alpha = 0.2, \alpha_{\text{red}}=0.95$ 
		\State $N_{\text{min}}=150, ~ ng = 10, ~ r_{\max}=0.95$
		\State $\tbf{Initialize tomogram: } \tbf{x} = \tbf{x}_0 = 0$
		\While{$k < N_{\text{proj}}$} \textit{(Dynamic Tilt Series Loop)}
		\State $\beta = \beta_0 \cdot (1 - \frac{5}{6} \cdot \frac{k}{N_{\text{proj}}})$
		\For {$j=1, N_{\text{min}}$} \textit{(ASD-POCS Loop)} \label{alg:asd_start}
		\State $\tbf{x}_0 = \tbf{x}$ 
		\State $\tbf{x} = \text{randART}(\tbf{x}, \beta, k)$
		\State $\tbf{x} = \max(0, \tbf{x})$
		\State $\tbf{g} = A\tbf{x}$ 
		\If {$k = 1 \tbf{~and~} j = 1 $}  
			\State $\alpha_0 = \alpha \cdot \| \tbf{x} - \tbf{x}_0 \|_2$
		\ElsIf {$k \neq 1 \tbf{~and~} j = 1  $}
			\State $\alpha = \alpha_0 $
		\EndIf
		\State $dp = \| \tbf{x} - \tbf{x}_0 \|_2$
		\State $dd = \| \tbf{b} - \tbf{g} \|_2 $
		\State $\tbf{x}_0 = \tbf{x}$
		\State $\tbf{x} = \text{tv\_gd\_3D}(\tbf{x}, \alpha, ng)$
		\State $dg =  \| \tbf{b} - \tbf{g} \|_2 $
		\If {$ dg > dp \cdot r_{\max} \tbf{~and~} dd > \epsilon $} 
			\State $ \alpha = \alpha_{\text{red}} \cdot \alpha $
		\EndIf
		\EndFor \label{alg:asd_end}
		\EndWhile
        \State \tbf{Output:} $\tbf{x} \in \mathbb{R}^3 \rightarrow \text{3D Tomogram}$
	\end{algorithmic}
\end{algorithm} 

\begin{algorithm} [H]
	\caption{Randomized ART (randART)}
	\begin{algorithmic}[1]
		\State \tbf{Input:} $\tbf{x} \in \mathbb{R}^3 \rightarrow \text{3D Tomogram to be constructed}$ 
		\State \tbf{Input:} $\beta \rightarrow \text{ART Parameter}$
		\State \tbf{Input:} $N_{\text{sample}} \rightarrow \text{Number of Projections Collected}$ 
		\Function{randART}{$ \tbf{x}, \beta, N_{\text{sample}}$}
		\For{$i = 1, N_{\text{slice}}$}
		\For{$j = 1, N_{\text{sample}}$}
		\State $\tau = \text{sample}(1 \dots, N_{\text{sample}})$
		\State $ \tbf{x}^{{i+1}} =\tbf{x}^{(i)} +  \beta a_{\tau}^T  \frac{b_{i,\tau} - \langle a_{\tau} , \tbf{x}^{(i)} \rangle }{\| a_{\tau} \| ^2} $
		\EndFor
		\EndFor
		\State \tbf{Output:} $\tbf{x} \in \mathbb{R}^3 \rightarrow \text{Current Iterate of 3D Tomogram}$
		\EndFunction
	\end{algorithmic}
\end{algorithm}

\newpage
\setlength{\intextsep}{5pt}
\thispagestyle{plain}
\begin{algorithm} [H]
	\caption{3D TV Minimization (TV\_GD\_3D)}
	\begin{algorithmic}[1]
		\State \tbf{Input:} $\tbf{x} \in \mathbb{R}^3 \rightarrow \text{3D Tomogram to be constructed}$ 
		\State \tbf{Input:} $\alpha \rightarrow \text{TV Parameter}$
		\State \tbf{Input:} $ng \rightarrow \text{Number of TV Iterations}$ 
		\Function{tv\_gd\_3D}{$ \tbf{x}, \alpha, ng$}
		\State $\sigma = 10^{-6}$
		\For{$i = 1, ng$}
		\For{$j = 1,\text{nx}$}
		\For{$k  = 1,\text{ny}$}
		\For{$l  = 1,\text{nz}$}
		\State $v_1 = \frac{3\cdot x_{i,j,k} - x_{i+1,j,k} - x_{i,j+1,k} - x_{i,j,k+1}  }{\sqrt{ \sigma+ (x_{i,j,k} - x_{i+1,j,k})^2 +  (x_{i,j,k} - x_{i,j+1,k})^2 + (x_{i,j,k} - x_{i,j,k+1})^2 }}$ 
		\State $v_2 = \frac{x_{i,j,k} - x_{i-1,j,k} }{\sqrt{ \sigma+ (x_{i-1,j,k} - x_{i,j,k})^2 +  (x_{i-1,j,k} - x_{i-1,j+1,k})^2 + (x_{i-1,j,k} - x_{i-1,j,k+1})^2 }}$ 
		\State $v_3 = \frac{x_{i,j,k} - x_{i,j-1,k} }{\sqrt{ \sigma+ (x_{i,j-1,k} - x_{i+1,j-1,k})^2 +  (x_{i,j-1,k} - x_{i,j,k})^2 + (x_{i,j-1,k} - x_{i-1,j-1,k+1})^2 }}$ 
		\State $v_4 = \frac{x_{i,j,k} - x_{i,j,k-1} }{\sqrt{ \sigma+ (x_{i,j,k-1} - x_{i+1,j,k-1})^2 +  (x_{i,j,k-1} - x_{i,j+1,k-1})^2 + (x_{i,j,k-1} - x_{i,j,k})^2 }}$   
		\State $\nabla_{\text{TV}}\tbf{x}_{j,k,l}  = v_1 + v_2 + v_3 + v_4 $
		\EndFor
		\EndFor
		\EndFor
		\State $\tbf{x}^{(i+1)} = \tbf{x}^{(i)} - \alpha \cdot \frac{ \nabla_{\text{TV}}\tbf{x} } {\|\nabla_{\text{TV}}\tbf{x}\|_2  }$
		\EndFor
		\State \tbf{Output:} $\tbf{x} \in \mathbb{R}^3 \rightarrow \text{Current Iterate of 3D Tomogram}$
		\EndFunction
	\end{algorithmic}
\end{algorithm}
\thispagestyle{plain}
\FloatBarrier

\pagebreak

\section{Supplemental Movies}
\thispagestyle{plain}
\begin{figure*}[ht]
    \includegraphics[width=0.9\linewidth]{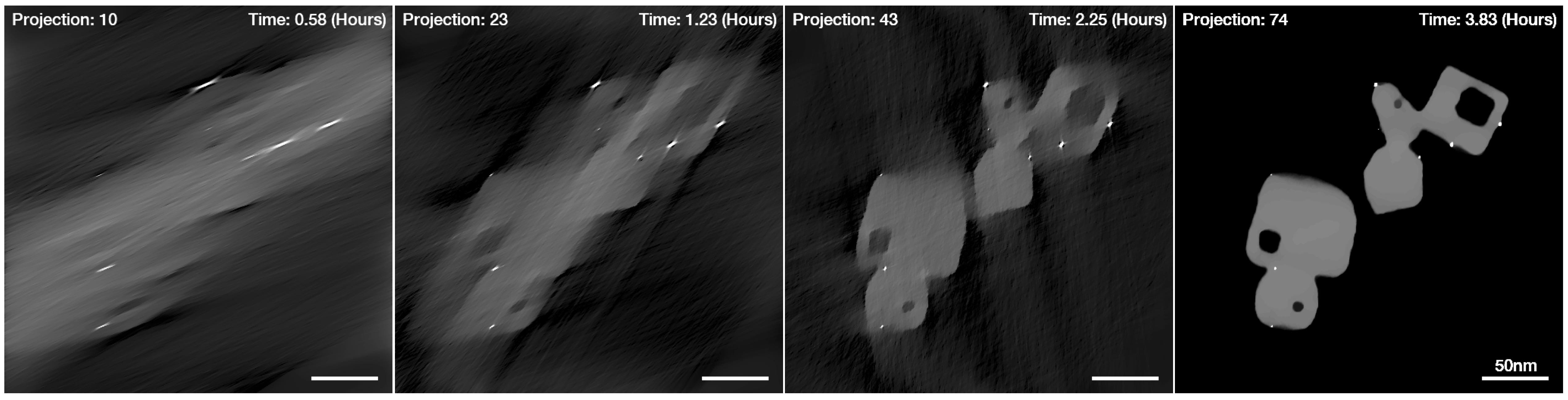}
    \caption{\textbf{Screenshots of Dynamic a Au/STO$_3$ Reconstruction from Supplemental Video 1.} 2D Slices of the 3D volume, illustrated in Figure 1, as the reconstruction progresses with increasing projections number (top-left) and time elapsed (top-right). }
    \label{fig:recon_movie_1}
\end{figure*}

\begin{figure*}[ht]
    \includegraphics[width=0.9\linewidth]{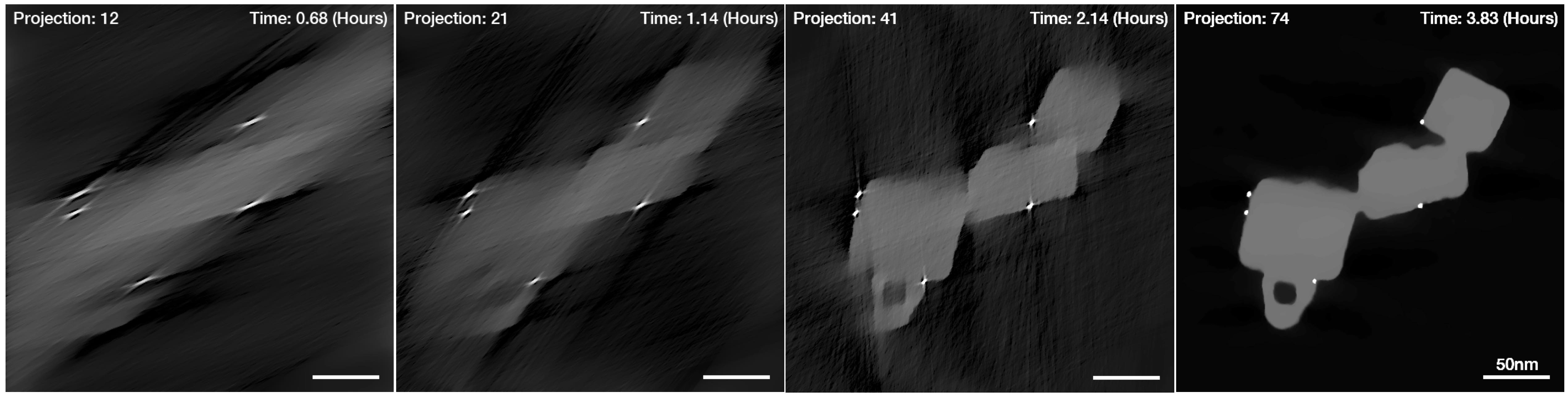}
    \caption{\textbf{Screenshots of Dynamic a Au/STO$_3$ Reconstruction from Supplemental Movie 2.} 2D Slices of the 3D volume, illustrated in Figure 2, as the reconstruction progresses with increasing projections number (top-left) and time elapsed (top-right).}
    \label{fig:recon_movie_2}
\end{figure*}

\begin{figure*}[ht]
    \includegraphics[width=0.9\linewidth]{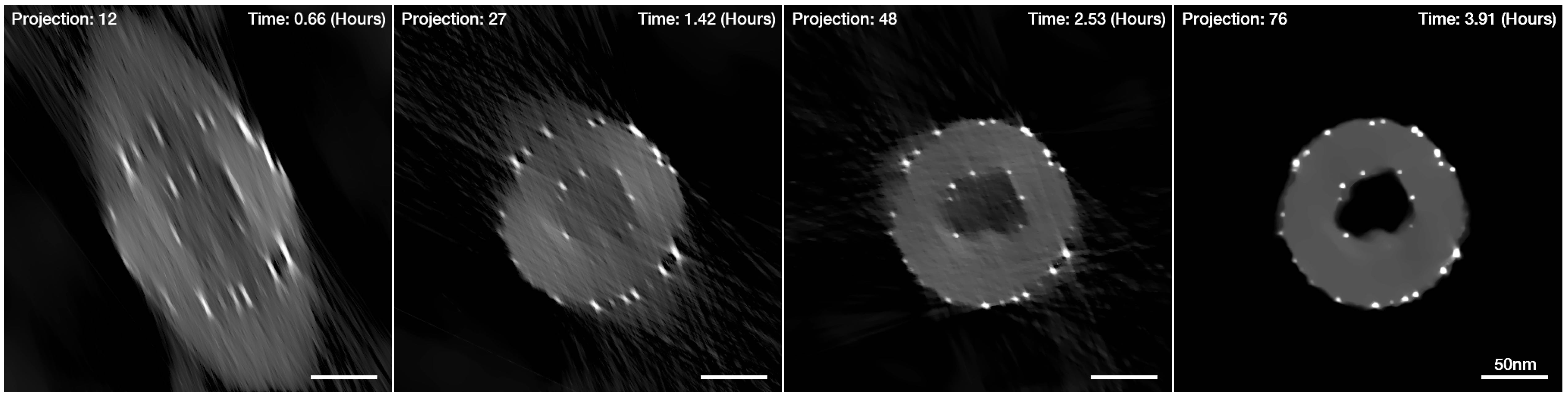}
    \caption{\textbf{Screenshots of Dynamic a Pt/Carbon Nanofiber Reconstruction from Supplemental Movie 3.} 2D Slices of the 3D volume, illustrated in Figure 3, as the reconstruction progresses with increasing projections number (top-left) and time elapsed (top-right).}
    \label{fig:recon_movie_3}
\end{figure*}

\end{document}